\documentclass[preprint,12pt]{elsarticle}

\usepackage{framed,multirow}

\usepackage{amssymb}
\usepackage{latexsym}
\usepackage{bm}
 \usepackage{indentfirst}
\usepackage{amsmath,amsthm,amssymb}

\usepackage{graphicx}

\usepackage{float}
\usepackage[framemethod=tikz]{mdframed}
\usepackage{epstopdf}
\usepackage{natbib}
\usepackage[a4paper, total={6.5in, 9in}]{geometry}
\usepackage{xcolor}
\usepackage{soul}
\usepackage{bm}
\usepackage{amsmath}
\usepackage{xcolor}

\usepackage{comment}

\usepackage{subcaption}

\usepackage{import}

\newif\ifturb

\newif\ifsutherland

\usepackage{url}
\usepackage{xcolor}
\definecolor{newcolor}{rgb}{.8,.349,.1}

\begin{document}


\begin{frontmatter}
\title{Towards a robust detection of viscous and turbulent flow regions using unsupervised Machine Learning}

\author[1]{Kheir-Eddine  Otmani\corref{cor1}}
\ead{otmani.kheir-eddine@alumnos.upm.es}
\author[1]{Gerasimos  Ntoukas}
\author[1,2]{Esteban Ferrer}
\cortext[cor1]{Corresponding author}
\address[1]{ETSIAE-UPM-School of Aeronautics, Universidad Politécnica de Madrid, Plaza Cardenal Cisneros 3, E-28040 Madrid, Spain}
\address[2]{Center for Computational Simulation, Universidad Politécnica de Madrid, Campus de Montegancedo, Boadilla del Monte, 28660 Madrid, Spain}


\begin{abstract}
We propose an invariant feature space for the detection of viscous dominated and turbulent regions (i.e., boundary layers and wakes). The developed methodology uses the principal invariants of the strain and rotational rate tensors as input to an unsupervised Machine Learning Gaussian mixture model. The selected feature space is independent of the coordinate frame used to generate the processed data, as it relies on the principal invariants of strain and rotational rate, which are Galilean invariants. This methodology allows us to identify two distinct flow regions: a viscous dominated, rotational region (boundary layer and wake region) and an inviscid, irrotational region (outer flow region). 
We test the methodology on a laminar and a turbulent (using Large Eddy Simulation) case for flows past a circular cylinder at $Re=40$ and $Re=3900$. The simulations have been conducted using a high-order nodal Discontinuous Galerkin Spectral Element Method (DGSEM). The results obtained are analysed to show that Gaussian mixture clustering provides an effective identification method of viscous dominated and rotational regions in the flow. We also include comparisons with traditional sensors to show that the proposed clustering does not depend on the selection of an arbitrary threshold, as required when using traditional sensors.

\end{abstract}
\end{frontmatter}

\tableofcontents

\section{Introduction}
The large amount of data generated by computational fluid dynamics (CFD) simulations has led scientists towards the use of new Machine Learning (ML) techniques to help perform post-processing analysis, such as the detection of flow regions or flow features \cite{computation8010015,brunton_2020}. These regions or features can subsequently be used to identify physical mechanisms or construct surrogate models. Alternatively, the regions can be used from a numerical perspective to increase local resolution (e.g. refine the mesh or increase the polynomial order in high-order methods) or to apply different sets of equations in each region (e.g. near wall turbulent region and inviscid far field). Overall, there is great potential in using Machine Learning techniques to identify flow regions, which is the main topic of this work.
To detect regions using machine learning, one can use classification or clustering \cite{AizedAminSoofiArshadAwan2017}. Both allow for the grouping of data by creating clusters of points/elements with similar properties/values. The fundamental problem to be solved is the determination of the boundary of these nuclei, which will differentiate the classified/clustered data. Classification algorithms (e.g., Logistic Regression, K-Nearest Neighbours, Decision Tree, Support Vector Machines) belong to the supervised learning category and require predetermined labelled data sets to guide and train the machine learning algorithm. Clustering (e.g., Gaussian mixture models, K-means, Mini-Batch K-Means, Mean Shift, Spectral Clustering) \cite{bishop_2006}, is an unsupervised learning method that does not require pre-sampled data to cluster regions and can automatically discover grouping in data. In this work, we select clustering as it requires minimum supervision.

The combination of fluid dynamics with neural networks and deep learning is an emerging field \cite{brunton_2020,vinuesa_brunton}.
Recently, deep neural networks have been used to build surrogates of closure turbulence models \cite{ling_2016,Tracey2015AML,zhang_2018,Geneva2019QuantifyingMF,Maulik2019AcceleratingRT}, to accelerate numerical simulations (e.g., \cite{bar_sinai_2019,fernando_2022,fernando2}) and to solve partial differential equations (e.g., \cite{SIRIGNANO20181339,RAISSI2019686,Weinan2017TheDR,Ranade2021DiscretizationNetAM}).

More relevant to this work is the recent use of ML for flow region identification in \cite{binglin_2020} as a supervised ML classification (supervised learning) method was used to identify turbulent and non-turbulent regions in a flow past a circular cylinder at a variety of turbulent Reynolds numbers. The authors used a variety of inputs (e.g., kinetic energy, vorticity) and invariants of strain rate and vortical tensors to train an extreme gradient boosting XGBoost classifier. 
In the context of the identification of turbulent/non-turbulent regions, \cite{wu_som_2019} used self-organising map clustering \cite{kohonen_som} to distinguish the turbulent boundary layer from non-turbulent regions in a transitional flow. These results were encouraging and motivated our work, which is based on the clustering method.
 The main advantage of using ML to perform flow region classification or clustering is that the methodology does not require the choice of any threshold and can treat multiple inputs. On the contrary, classic detectors heavily dependent on the selected threshold, as first pointed out by \cite{wu_som_2019}, and corroborated in section \ref{sec:results}.
 
When treating Reynolds Averaged Navier-Stokes (RANS) simulations, \cite{ettore_2022,lanzetta_2015} showed that various viscous sensors could be used as input to a clustering ML framework. The aim was to detect the boundary layer and wake regions for turbulent flow past a NACA 0012 airfoil. The methodology combined viscous sensors to perform a soft clustering of the given data. The results showed that the ML clustering outperformed the classic viscous sensors used to identify the flow regions in RANS. In
\cite{callaham_2021}, unsupervised learning techniques were proposed to identify the dominant physical processes for different flow scenarios and in \cite{colvert_2017} the authors trained a neural network to classify different types of vortex wakes. Authors in \cite{BECK2020109824} used supervised learning and convolution neural networks \cite{Alzubaidi2021ReviewOD} to locate shock positions within a discontinuous Galerkin solver and used the output of this framework for shock capturing applications.

In relation to the parameter space for the clustering method, \cite{ling_2016} showed in the context of Reynolds averaged turbulence modelling that using Galilean invariants to train the ML model is beneficial since it ensures independence between the coordinate inertial frame used to generate the data and the predictions made by the ML model. 

In this work, we aim at distinguishing not only turbulent/non-turbulent regions but also at detecting viscous dominated regions from outer inviscid regions. To do so, we construct a robust flow feature space (i.e., useful for subsonic laminar and turbulent regimes) and allow for two clustering regions. The first is a viscous and turbulent dominated region and the second is an inviscid outer region. Note that this is different from \cite{wu_som_2019}, where turbulent and laminar regions are distinct clusters.
We aim to distinguish near wall and wake regions from inviscid/potential flow regions (far from objects in the flow). We propose to use a data-driven clustering approach, the Gaussian mixture clustering. The selection of the feature space, which is fed into the Gaussian mixture clustering is critical and has to carry the physical information needed to distinguish the boundary layer and wake regions from the outer flow region, see details in section \ref{sec:meth}. Such a space is constrained to maintain the Galilean invariance. To test our proposed methodology, we conduct numerical experiments on a laminar and a turbulent flow past a circular cylinder.

The reason for having only two regions (laminar/turbulent and inviscid) is that an extension of this work will be to locally increase the resolution in these regions (e.g. refine the mesh or increase the polynomial order in high-order methods) to improve the accuracy in our simulations. Alternatively, these regions can be used to compute drag, as proposed for RANS in \cite{ettore_2022,lanzetta_2015}.

The rest of this work is structured as follows. First, we introduce the ML methodology and the selected flow features to train the ML model in section~\ref{sec:meth} and denote the details of the numerical experiments for the laminar and turbulent (LES) cases. In section~\ref{sec:results}, we present the results along with the analysis of the clustering and the resulting flow regions.

\section{Methodology}\label{sec:meth}
We propose a ML framework to detect viscous and turbulent (i.e., highly rotational) flow regions. 
The methodology is based on the assumption that the boundary and wake regions are characterized by non-negligible viscous dissipation and high vorticity/rotation in the flow field while the outer flow region is inviscid and irrotational (e.g., potential flow). Note that this will be verified a-posteriori in our analysis, as presented in section~\ref{sec:results}. We construct a robust flow feature space to be used as input to an unsupervised ML framework. Intuitively, this feature space must be independent of the coordinate frame used to generate the data and for this reason we propose to use the principal invariants of strain rate and rotational rate tensors.
The two strain rate tensor invariants are defined as 
\begin{align*} 
{ Q_{S}}=\frac{1}{2}( tr(\boldsymbol {  S})^2-tr(\boldsymbol { S}^2))\;\; ; \;\;
 R_{S}=-\frac{1}{3}det(\boldsymbol { S}),
\end{align*}
where $\boldsymbol{  S}$ is the strain rate tensor defined as: $  \boldsymbol {  S}=\frac{1}{2}(\boldsymbol { J}+\boldsymbol { J}^T)$ and $ \boldsymbol{ J}=\nabla\boldsymbol{  U}$ is the gradient tensor of the velocity field $\boldsymbol{  U}$. $ Q_{S}$ is proportional to the local viscous dissipation rate $\epsilon=-4\mu Q_{S}$, with $\mu$ denoting the fluid viscosity \cite{zhou_2015}. $R_{S}$ is relevant as it relates to regions of high viscous dissipation. Positive values of $R_{S}$ indicate high rates of strain production, while negative values of $R_{S}$ indicate the destruction of the strain product \cite{da_silva_2008}. The rotational tensor $\boldsymbol { \Omega}$ has only one invariant defined as 
\begin{align*}
    Q_{\Omega}=-\frac{1}{2}tr( \boldsymbol { \Omega}^2),
\end{align*}
where $\boldsymbol {  \Omega}=\frac{1}{2}(\boldsymbol { J}-\boldsymbol {  J}^T)$. $ Q_{\boldsymbol \Omega}$ is related to the enstrophy density $\boldsymbol{\xi}$ \cite{zhou_2015} defined by
$\boldsymbol \xi= \int_D |\boldsymbol \omega|^2 dD, $
where $\boldsymbol \omega$ is the flow vorticity in the domain $D$. Enstrophy can be associated to high turbulent dissipation rate and can be used to detect turbulent regions. 
High values of $Q_{\Omega}$ identify rotational and turbulent regions in the flow field.
Using these quantities as input, we propose the feature space 
\begin{align*}
    E=\left( Q_{S},R_{S}, Q_{\Omega} \right).
\end{align*}
This feature space $E$ is used as input to the Gaussian mixture model to cluster the data into two different regions. The boundary layer and wake regions (viscous and turbulent dominated) and the outer flow region (inviscid and irrotational). The Gaussian mixture model is preferred over simpler clustering (e.g. K-means) for its superiority in discovering complex non-linear patterns in the data \cite{gmmvsk-means_2020}. 
We will test this feature space with two different regimes (laminar $Re=40$ and turbulent $Re=3900$) to verify that we can distinguish laminar/turbulent dominated regions from outer inviscid regions.
In section \ref{sec:results}, we demonstrate why these 3 invariants are sufficient to detect the two regions of interest. Furthermore, we will showcase that it is not advantageous to reduce the feature space as all three invariants are necessary.


\subsection{Gaussian mixture}\label{sec:clustering}
Clustering is the process of categorising data into different groups. This is essentially performed by discovering underlying patterns within the given data \cite{saxena_2017}. It should be noted that clustering is an unsupervised learning approach which can separate data without the necessity of specifying a ground truth, unlike in supervised approaches, which require expert guidance by providing a set of labels to guide the learning process. 

 The model adopted in this work is the Gaussian Mixture Model (GMM) \cite{bishop_2006}.  
Under the hypothesis that the data are generated from a mixture of Gaussian distributions, GMM clusters the data into different subpopulations, each following a Gaussian distribution. The algorithm estimates the mean and variance of each normal distribution iteratively using the Expectation-Maximization (EM) method \cite{Dempster_1977} to provide the optimal estimation of these parameters. The EM method consists of two main steps. The expectation step (E-step), in which the conditional expectation of the data is computed, given the available samples. This is done essentially to fill in the missing values in the given data. The second step is the maximisation step (M-step), where the maximisation of the conditional expectation, computed previously in the E-step, is used to update the mean and variance of the normal distributions. These two steps are repeated iteratively until no significant changes are observed in the estimated parameters. Namely, we use a tolerance of $10^{-4}$ to stop the algorithm with regard to the difference of the conditional probabilities from the previous iteration \cite{mclachlan_2019}. 

For clustering purposes, GMM considers each Gaussian distribution as a cluster and assigns each data sample to a cluster based on a membership probability. The number of Gaussian distributions $N$ should be provided prior to the model training process. In this work, we select $N=2$ as we aim to detect two flow regions, a viscous-dominated, rotational region (boundary layer and wake region) and an inviscid irrotational region (outer flow region). Several clustering algorithms are available in open source libraries such as \textit{scikit-learn}  \cite{Pedregosa2011ScikitlearnML} which offers different choices based on different unsupervised learning frameworks. Our implementation uses the Gaussian mixture class in the \textit{scikit-learn} Python library \cite{Pedregosa2011ScikitlearnML}.

\subsection{Traditional sensors}\label{sec:trad_sensors}
In section \ref{sec:results}, the results from our proposed clustering method will be compared against two classic sensors used for the classification of flow regions, the dissipation of kinetic energy and the eddy viscosity sensor.

The first sensor under consideration is the dissipation of kinetic energy, which is defined as
\begin{eqnarray*}
      \frac{F_{\Phi}}{\mu} & = & 2\left(\left( \frac{\partial u}{\partial x}\right)^2+\left( \frac{\partial v}{\partial y}\right)^2+\left(\frac{\partial w}{\partial z}\right)^2\right)+ \left(\frac{\partial v}{\partial x} + \frac{\partial u}{\partial y}  \right)^2+\left(\frac{\partial w}{\partial y} + \frac{\partial v}{\partial z} \right)^2+  \left(\frac{\partial u}{\partial z} +  \frac{\partial w}{\partial x} \right)^2  \\
    &  &  -\frac{2}{3}  \left( \frac{\partial u}{\partial x}+ \frac{\partial v}{\partial y}+\frac{\partial w}{\partial z }\right) ,
\end{eqnarray*}
where $ u, v, w$ are the three velocity components \cite{schlichting_boundary-layer_2017}.

We also select a turbulent sensor: the eddy viscosity sensor \cite{paparone_2003} defined as 
\begin{align*}
    F_{\mu_t}=\frac{\mu+\mu_t}{\mu},
\end{align*}
where $ \mu$ and $\mu_{t}$ are the physical and turbulent viscosities accordingly. Note that this sensor detects turbulent regions ($F_{\mu_t}>1$ if $\mu_t>0$) and is unable to detect laminar regions within the boundary layer and the wake ( regions where $ \mu_t \to 0$ ) \cite{lanzetta_2015}.

To detect the boundary layer and wake regions using the aforementioned sensors, it is necessary to choose a threshold parameter $K$. Regions where $F_{\Phi}>K$ or $F_{\mu_t}>K$ will be considered as turbulent regions (boundary layer and wake regions), whereas regions where $F_{\Phi} \leq  K$ or $F_{\mu_t} \leq K$ will be considered as an inviscid, irrotational region (outer region). 
The identification of a suitable value for the threshold parameter $K$ is arbitrary and non-trivial \cite{lanzetta_2015} and is often determined through a trial and error process. We will show that our proposed clustering does not require any specification of a threshold parameter.

\subsection{High order discontinous Galerkin solver}
The numerical simulations have been carried out using the HORSES3D numerical framework, see \cite{horses_paper}. The compressible Navier-Stokes equations have been discretised using a high-order Discontinuous Galerkin Spectral Element Method (DGSEM) \cite{RuedaRamirez2021ASC}. The computational domain is tessellated into non-overlapping curvilinear hexahedral elements to approximate complex geometries. The solution within each element is represented with an arbitrary polynomial approximation order and can be discontinuous across different elements. The discontinuities between elements are treated through the use of suitable fluxes. In this work we use the Roe Riemann solver \cite{roe} for the convective fluxes and BR1 for the viscous fluxes \cite{br1}. The multiphysics environment of HORSES3D also offers the option of using several subgrid turbulence models. We use the Smagorinsky LES model for the turbulent cylinder at $Re=3900$ \cite{ GENERALCIRCULATIONEXPERIMENTSWITHTHEPRIMITIVEEQUATIONS}. Time marching is conducted using a low-storage $3^{rd}$ order explicit Runge-Kutta scheme \cite{WILLIAMSON198048}.
In all cases, the Mach number is 0.1 and the flow is considered incompressible. More details for each of the test cases are included in the following sections.

\subsection{Numerical Simulations}\label{sec:simulation}
We train the Gaussian mixture algorithm with the data obtained from numerical simulations of the flow past a circular cylinder in two different regimes: a laminar 2D steady flow at $Re=40$, a turbulent 3D flow at $Re=3900$, with a Smagorinsky subgrid closure model.

\subsubsection{Numerical simulation at $Re=40$}
\begin{figure}
     \centering
     \begin{subfigure}[H]{0.45\textwidth}
         \centering
         \includegraphics[width=\textwidth]{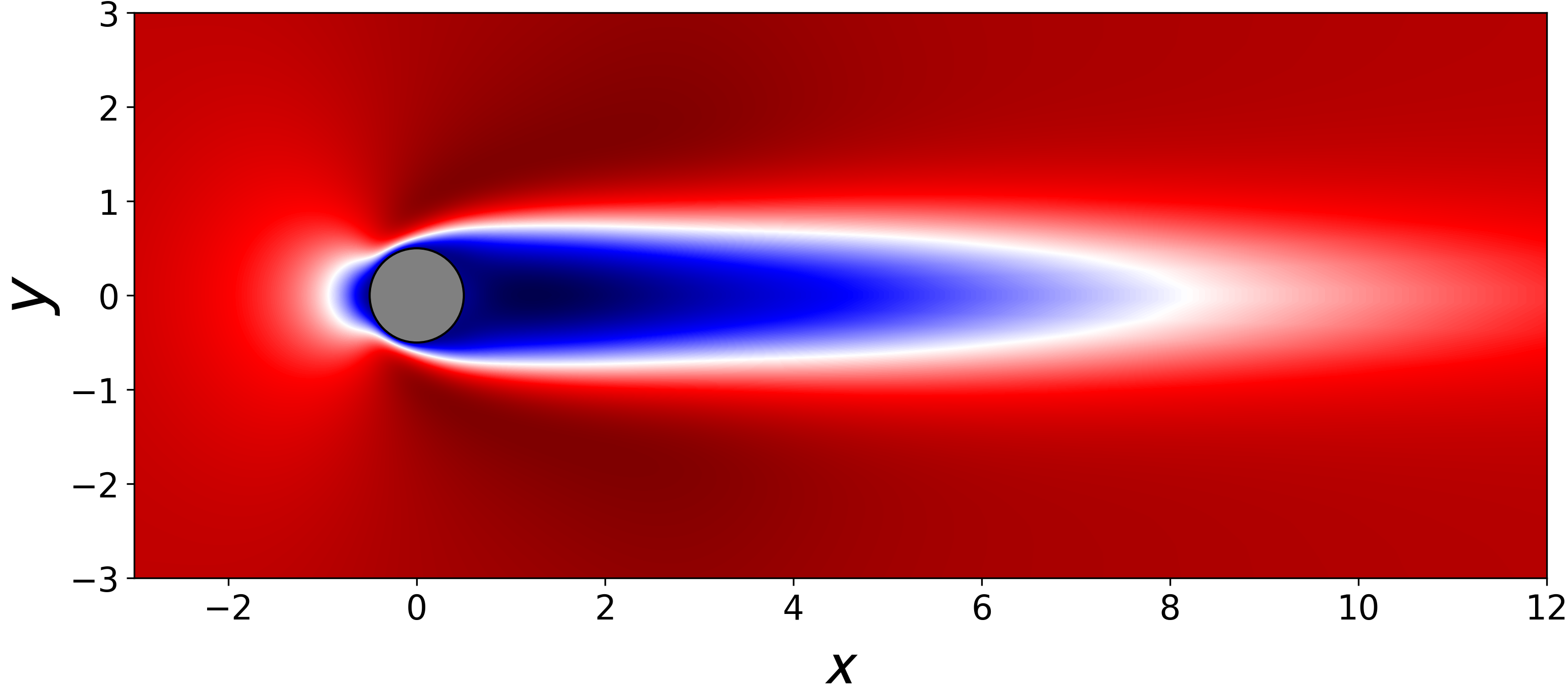}
         \caption{Streamwise velocity magnitude $\boldsymbol u$ for a flow past a cylinder at $Re=40$.}
         \label{u_40}
     \end{subfigure}
     \hfill
      \begin{subfigure}[H]{0.45\textwidth}
         \centering
         \includegraphics[width=\textwidth]{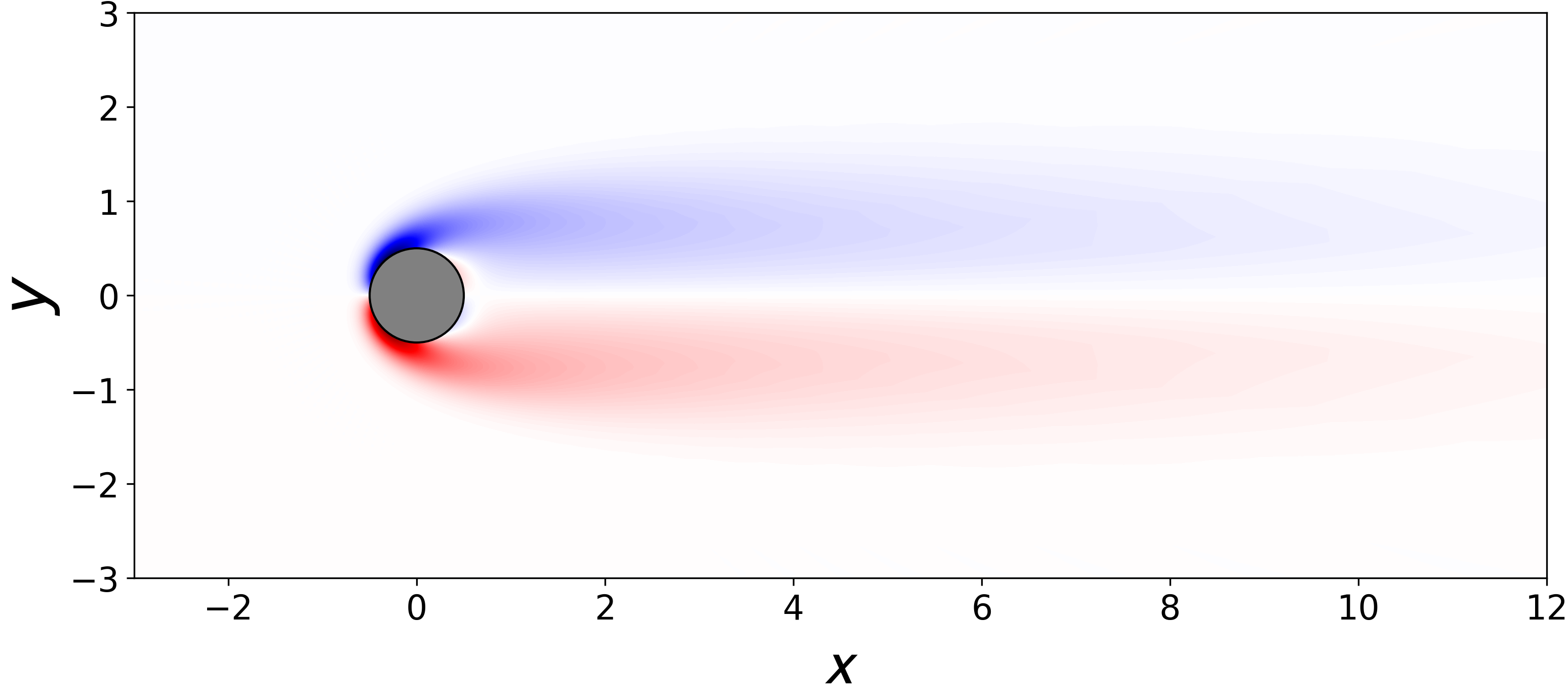}
         \caption{Spanwise vorticty component $\boldsymbol \omega_{z}$ for a flow past a cylinder at $Re=40$.}
         \label{Omega_z_40}
     \end{subfigure}
     \caption{Streamwise velocity $ u$ and spanwise vorticity $\omega_{z}$ for a flow past a circular cylinder at $Re=40$.}
     \label{u_omega_Re_40}
\end{figure}

We use a polynomial of order 2 (i.e., $3^{rd}$ order accurate) in 53088 hexahedral elements. Our methodology will cluster the degrees of freedom resulting from the high-order discretisation. Therefore, all nodal points provided by the polynomial order within each element will be treated by the clustering technique, with a total of $53088\times (P+1)^3 = 1.43 \times 10^6$ degrees of freedom.
Figure \ref{u_omega_Re_40} presents contours of the magnitude of the stream velocity $u$ and the spanwise vorticity component $ \omega_{z}$. 

\subsubsection{Numerical simulation at $Re=3900$}

\begin{figure}
     \centering
     \begin{subfigure}[htp]{0.45\textwidth}
         \centering
         \includegraphics[width=\textwidth]{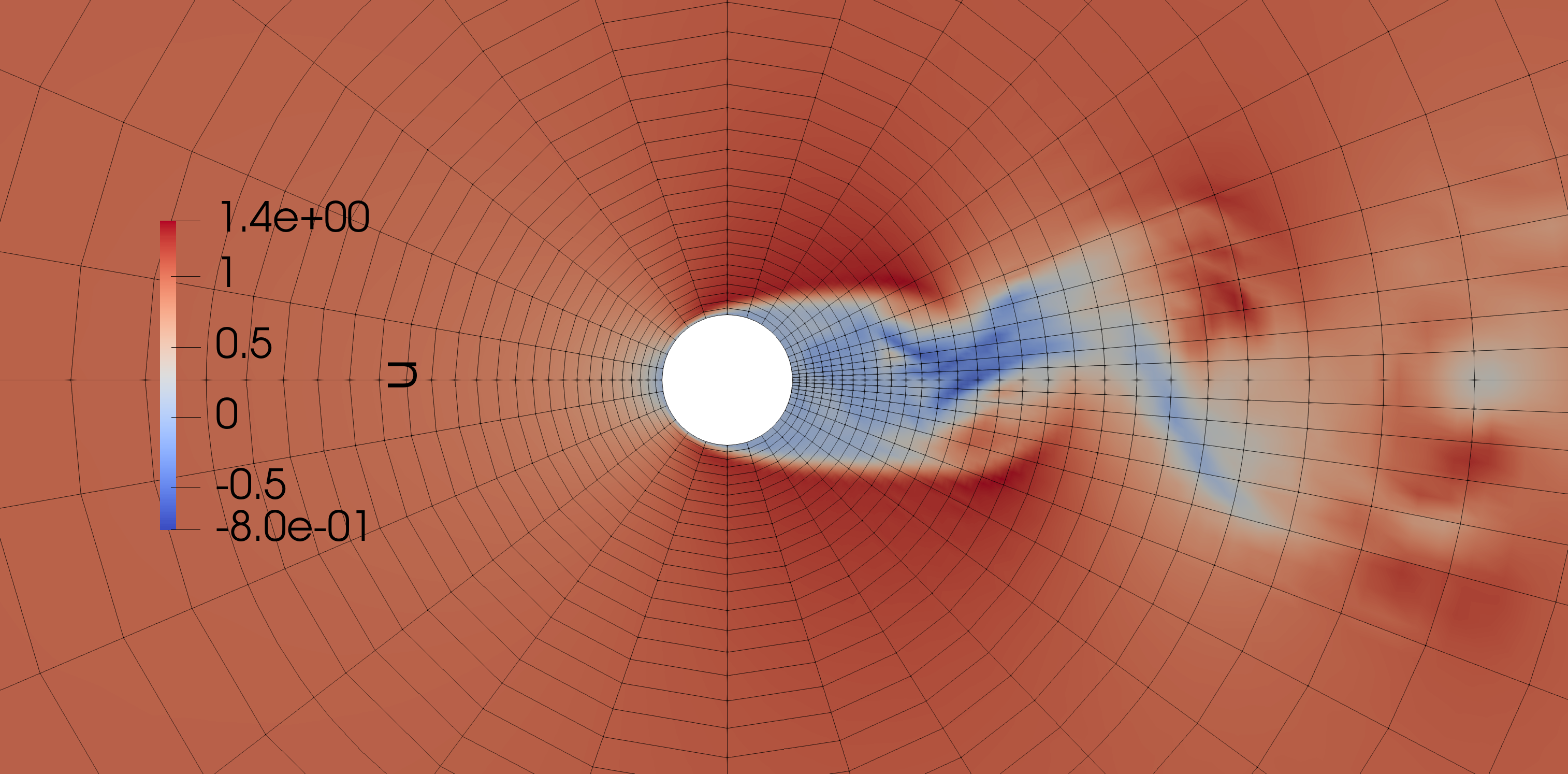}
         \caption{Close-up of the axial velocity $u$ along the midline and the mesh used for the flow around a cylinder at $Re_3900$.}
         \label{mesh_cyl3900}
     \end{subfigure}
     \hfill
      \begin{subfigure}[htp]{0.45\textwidth}
         \centering
         \includegraphics[width=\textwidth]{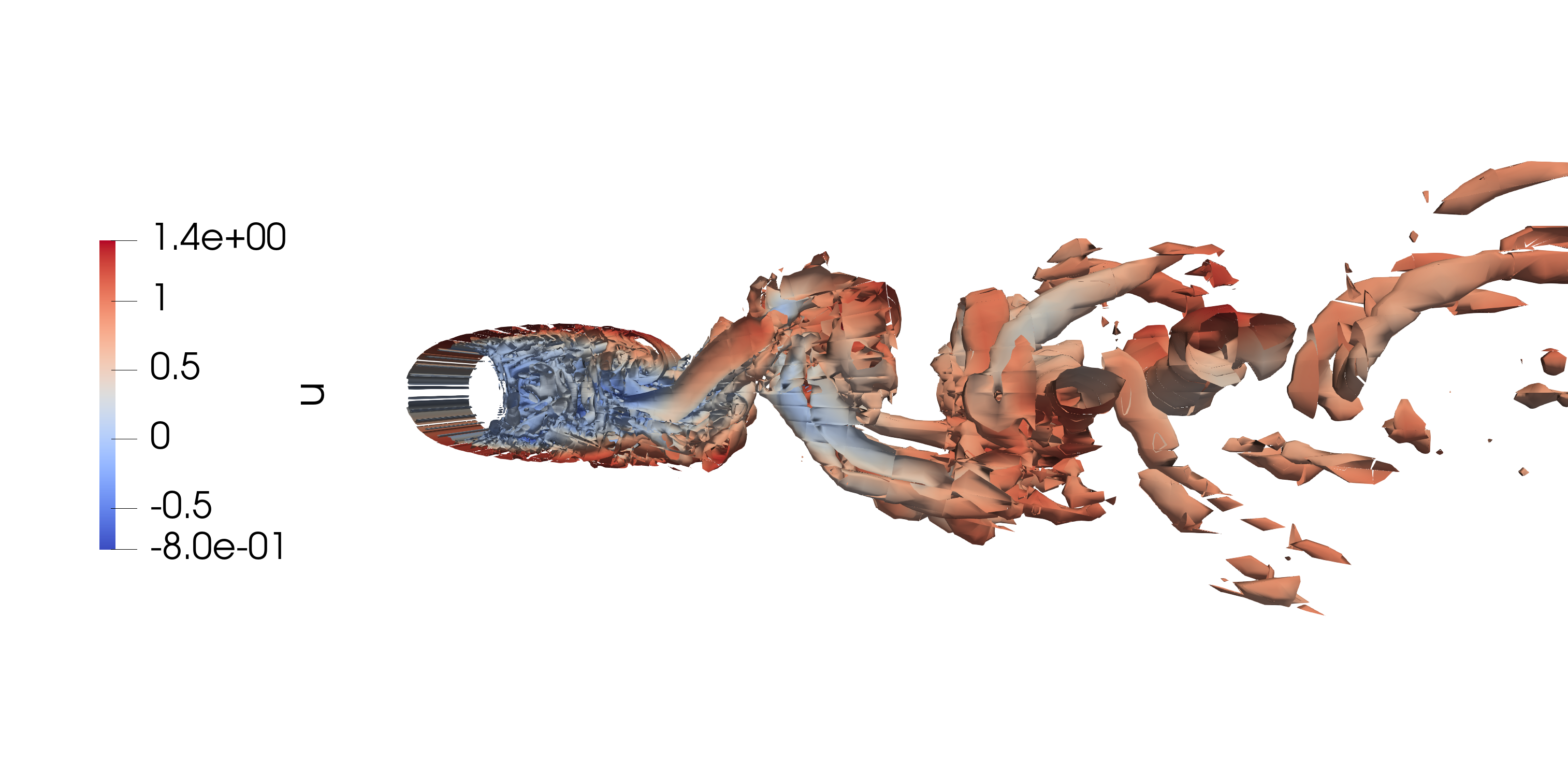}
         \caption{Isosurface of vorticty magnitude $||\boldsymbol\omega||=1$ coloured with the axial velocity $u$ for a flow past a cylinder at $Re=3900$.}
         \label{Omega_z_3900}
     \end{subfigure}
     \caption{Mesh and vorticity magnitude isosurface $||\boldsymbol\omega||$ for the flow past a circular cylinder at $Re=3900$.}
     \label{u_omega_Re_3900}
\end{figure}

\begin{figure}
     \centering
     \begin{subfigure}[htp]{0.49\textwidth}
         \centering
         \includegraphics[width=\textwidth]{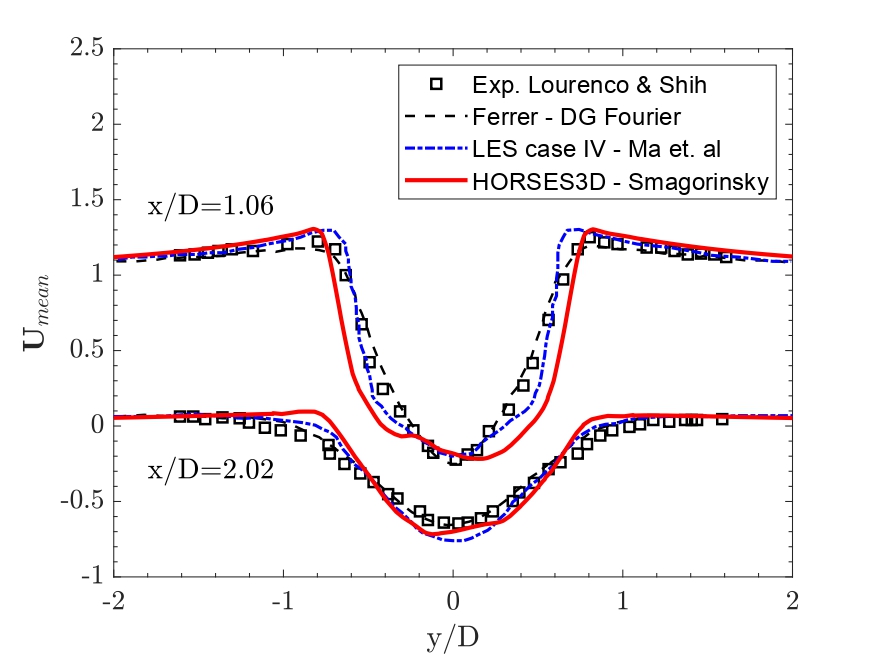}
         \caption{}
         \label{cyl3900_umean}
     \end{subfigure}
     \hfill
      \begin{subfigure}[htp]{0.49\textwidth}
         \centering
         \includegraphics[width=\textwidth]{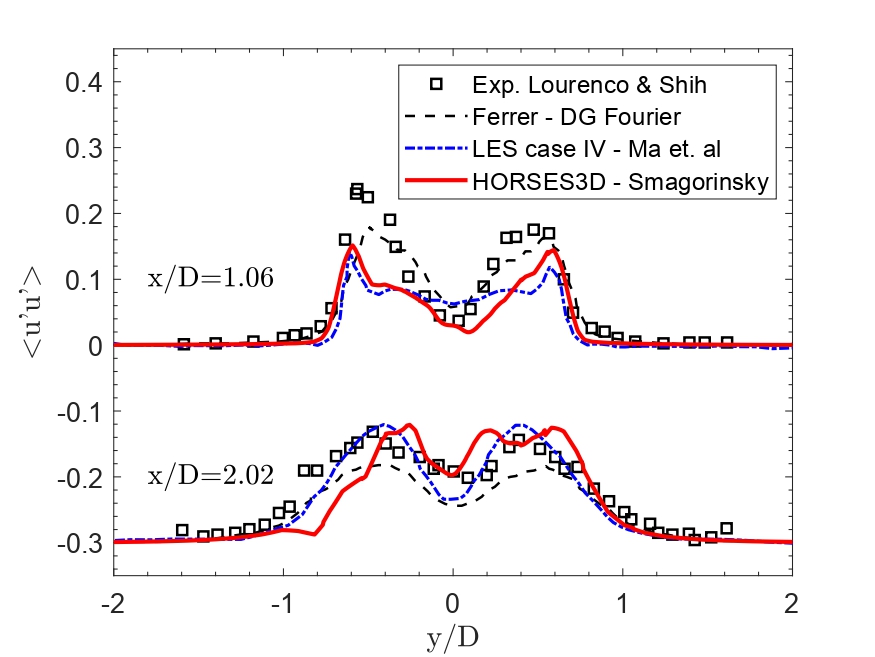}
         \caption{}
         \label{cyl3900_uu}
     \end{subfigure}
     \caption{Mean streamwise velocity $\textbf{{U}}_{mean}$ in (\ref{cyl3900_umean}) and streamwise Reynolds stresses {$<u'u'>$} in (\ref{cyl3900_uu}) downstream of the cylinder at locations $x/D=1.06$ and $x/D=2.02$ for the flow past a circular cylinder at $Re=3900$. We compare the data from HORSES3D against the data from Ferrer \cite{ferrer2017interior}, the LES results of Ma et al. \cite{ma2000dynamics} and the experimental data from Lourenco and Shih \cite{lourenco1994characteristics}.}
     \label{Cyl_Re_3900_stats}
\end{figure}

This case has been studied with various numerical frameworks and schemes \cite{ferrer2017interior,ma2000dynamics,parnaudeau2008experimental}. The flow has been computed using polynomial order 4 approximation (i.e., $5^{th}$ order) with a hexahedral mesh of 20736 elements, as shown in figure \ref{mesh_cyl3900}. The mesh has been extruded in the spanwise direction as $L_z/D=\pi$ and subdivided into 16 elements along this direction, with a polynomial of order 4 also used along this direction. In this case, the number of clustered points (degrees of freedom) is $20736\times (P+1)^3 = 2.59\times 10^6$.
 To ensure that aliasing errors are minimised and the method is robust, split form discretization with Pirozzoli averaging has been used \cite{pirozzoli2010generalized}. In figure~\ref{Cyl_Re_3900_stats}, we present a comparison for two different wake statistics quantities against other experimental and numerical results for this case.
 
%
%
\section{Results}\label{sec:results}
In this section, we present the results obtained using the Gaussian mixture algorithm to distinguish the boundary layer and wake region from the outer flow region. 

\subsection{Cylinder flow at Re=40}
\begin{figure}
\vspace{-5mm}
    \centering
     \begin{subfigure}[t]{0.45\textwidth}
     \includegraphics[width=\textwidth]{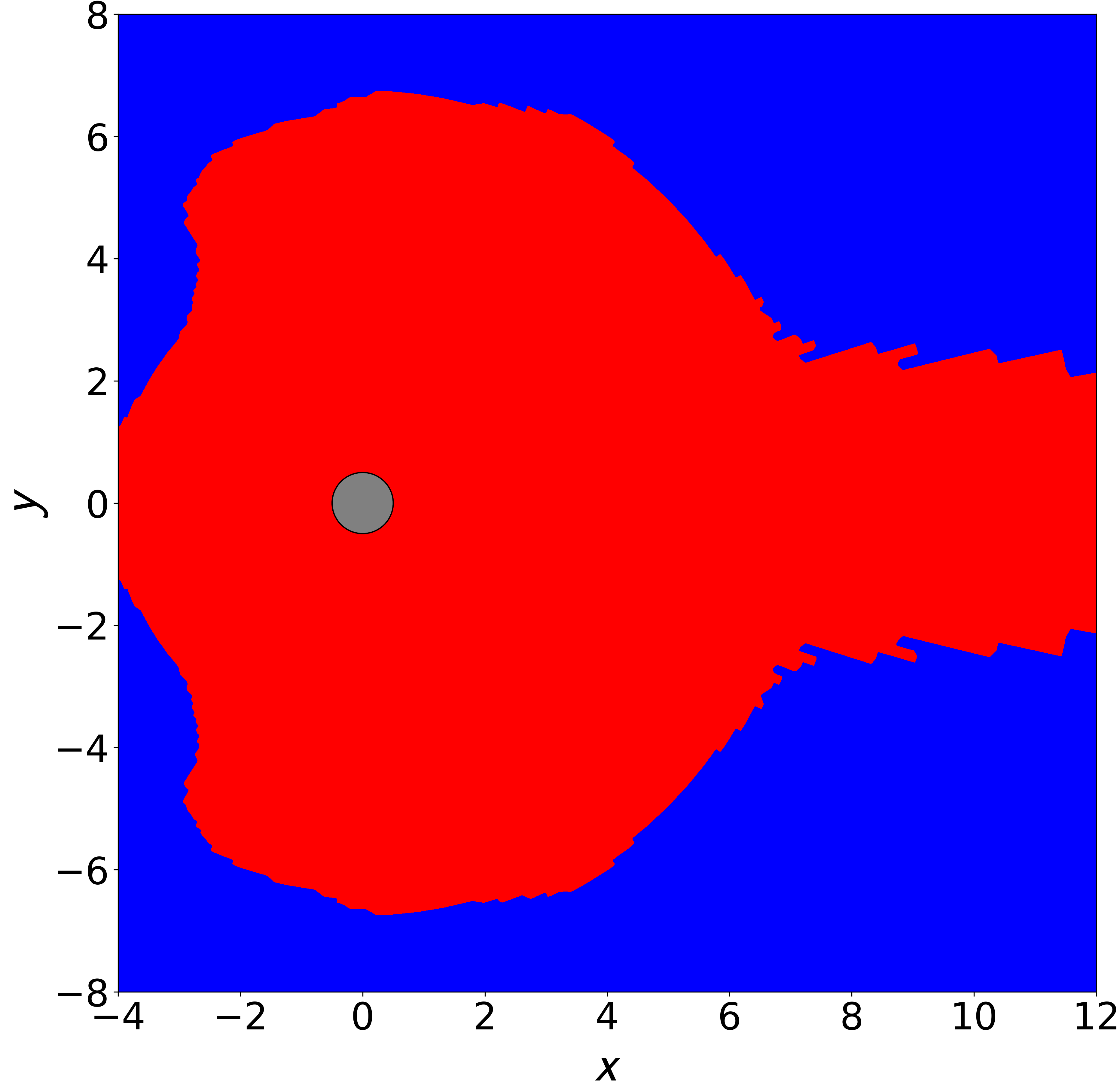}
     \caption{ }
    \label{regions_k_10}
    \end{subfigure}
    \hfill
      \begin{subfigure}[t]{0.45\textwidth}
     \includegraphics[width=\textwidth]{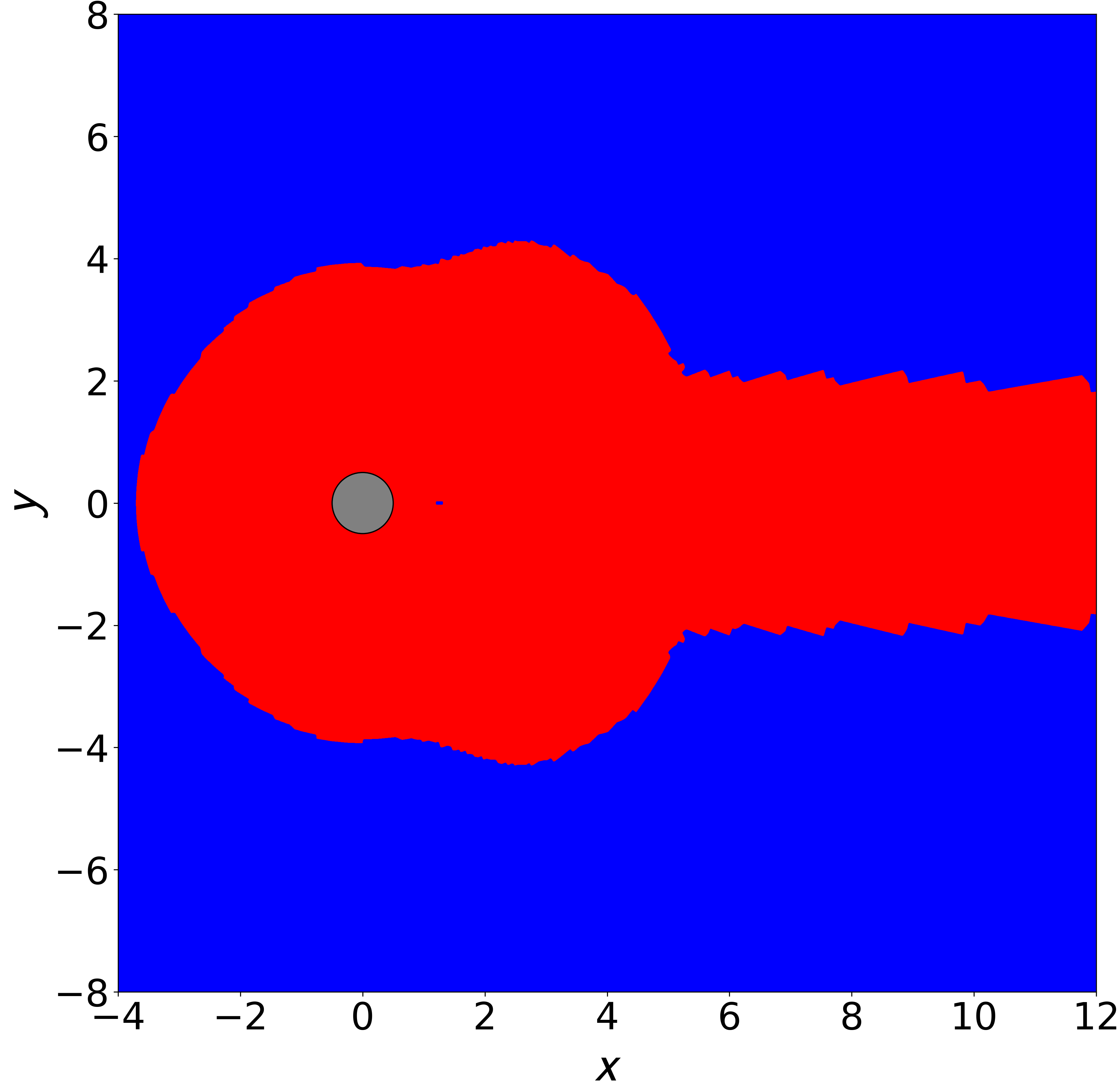}
    
    \caption{ }
    \label{regions_k_50}
    \end{subfigure}
    \hfill
     \begin{subfigure}[t]{0.45\textwidth}
      \includegraphics[width=\textwidth]{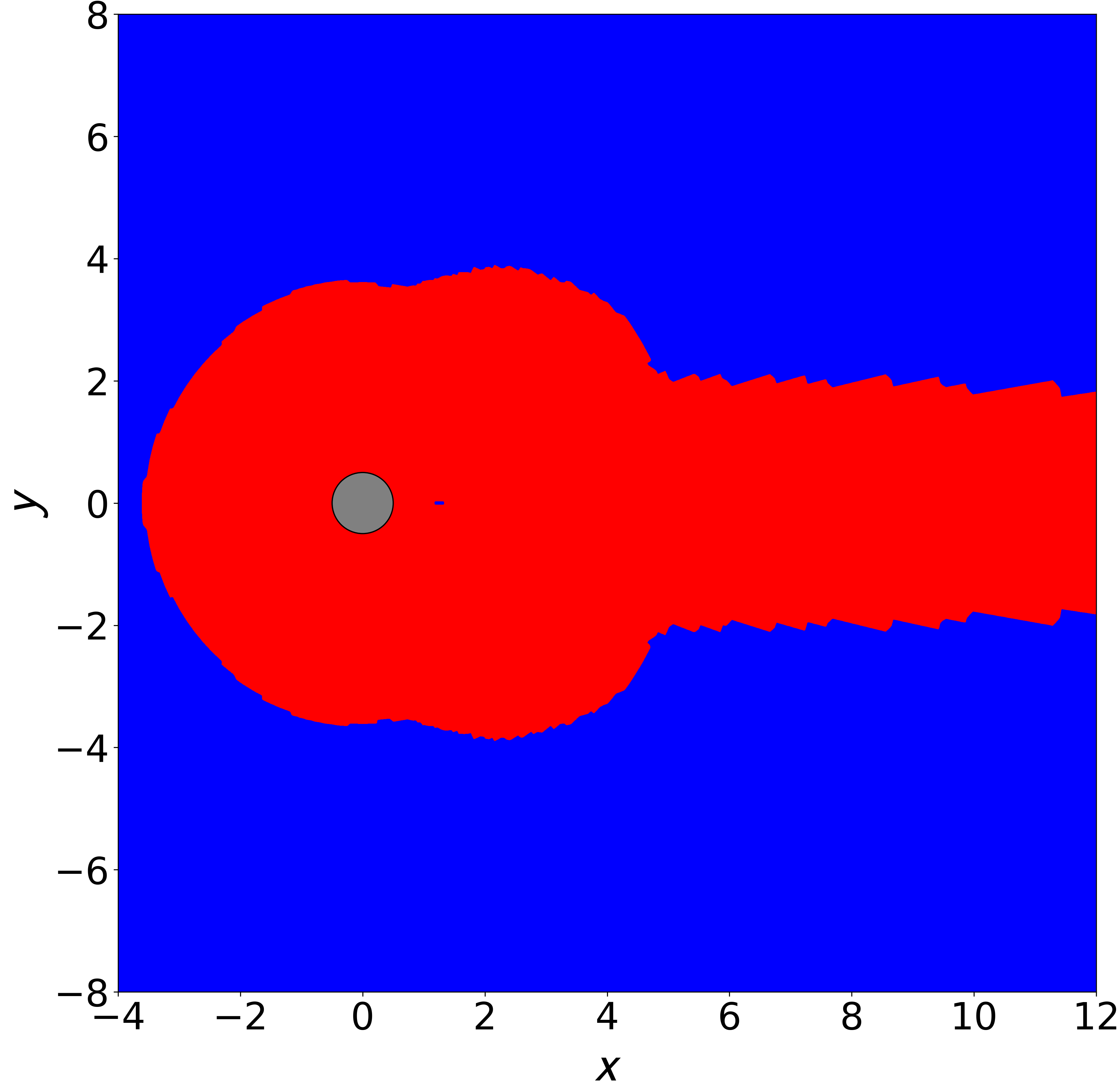}
      \caption{ }
      \label{regions_k_70}
     \end{subfigure}
    \hfill
    \begin{subfigure}[t]{0.45\textwidth}
     \includegraphics[width=\textwidth]{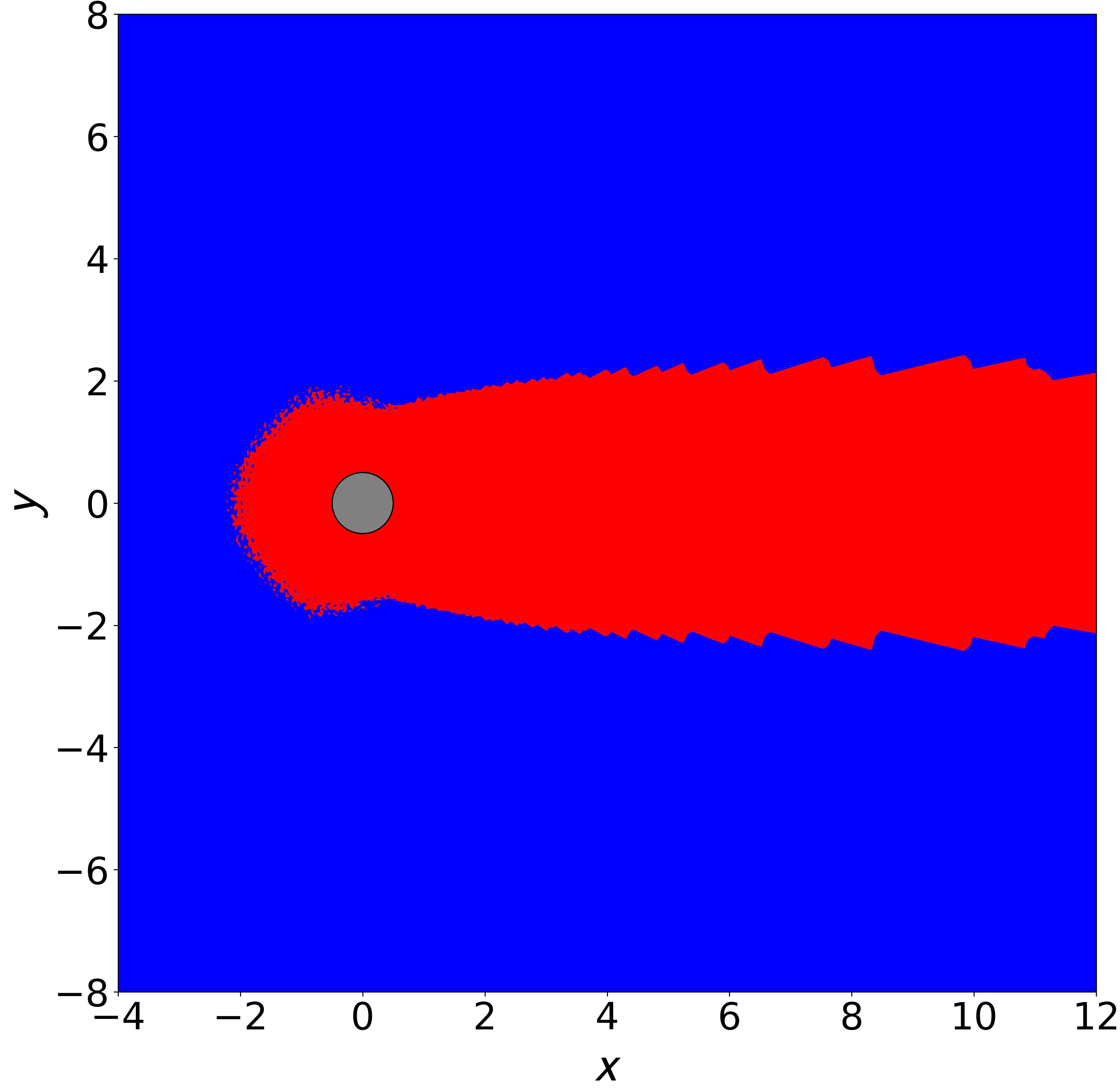}
     \caption{ }
     \label{GM_40}
    \end{subfigure}
    \caption{Flow regions detection for flow past a cylinder at $Re=40$ using dissipation of kinetic energy sensor $F_{\Phi}$, $K=10$ (\ref{regions_k_10}), $K=50$  (\ref{regions_k_50}), $K=70$ (\ref{regions_k_70}) and GMM clustering with feature space $E$. (\ref{GM_40}). \textcolor{red}{Red}: Boundary layer and wake regions, \textcolor{blue}{Blue}: Outer flow region.}
    \label{regions_re_40}
\end{figure}

To validate our methodology, we cluster the data obtained from a numerical simulation at $Re=40$ with the setup described in section \ref{sec:simulation}. Using the feature space $ E$ as input for the GMM, the detected regions are presented in figure \ref{GM_40}. The results of the GMM approach are compared in figure~\ref{regions_re_40} with the results obtained using the kinetic energy dissipation sensor $F_{\Phi}$, see section \ref{sec:trad_sensors}, with different values for the threshold parameter $K \in[10,70]$.
%
As presented in figures~\ref{regions_k_10}, \ref{regions_k_50}, \ref{regions_k_70}, the results with the traditional sensor are highly sensitive to the choice of the threshold parameter $K$ and different values of $K$ lead to very different regions and misidentify part of the outer region as part of the viscous dominated region. An appropriate choice of $K$ is challenging and requires a trial and error process, whereas ML clustering is free from selecting any threshold value.
Figure~\ref{regions_re_40} shows that our GMM clustering can provide a satisfactory detection of the boundary layer and wake regions.

\begin{figure}
     \centering
     \begin{subfigure}[h!]{0.45\textwidth}
         \centering
         \includegraphics[width=\textwidth]{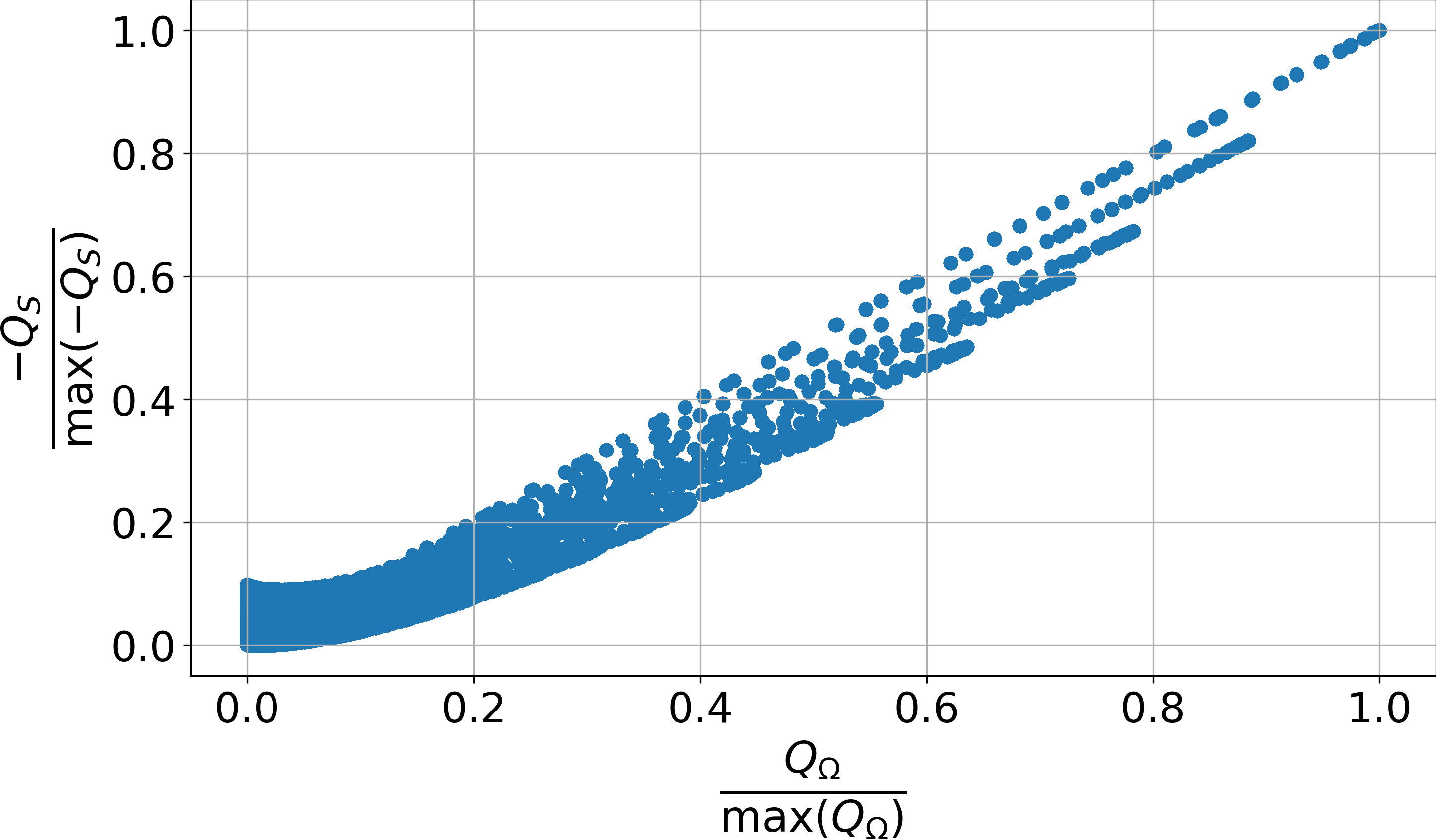}
         \caption{$(Q_\Omega,-Q_S)$ scatter plot in the detected boundary layer and wake region.}
         \label{Q_sVsQ_omega_viscous_Re=40}
     \end{subfigure}
     \hfill
     \begin{subfigure}[h!]{0.45\textwidth}
         \centering
         \includegraphics[width=\textwidth]{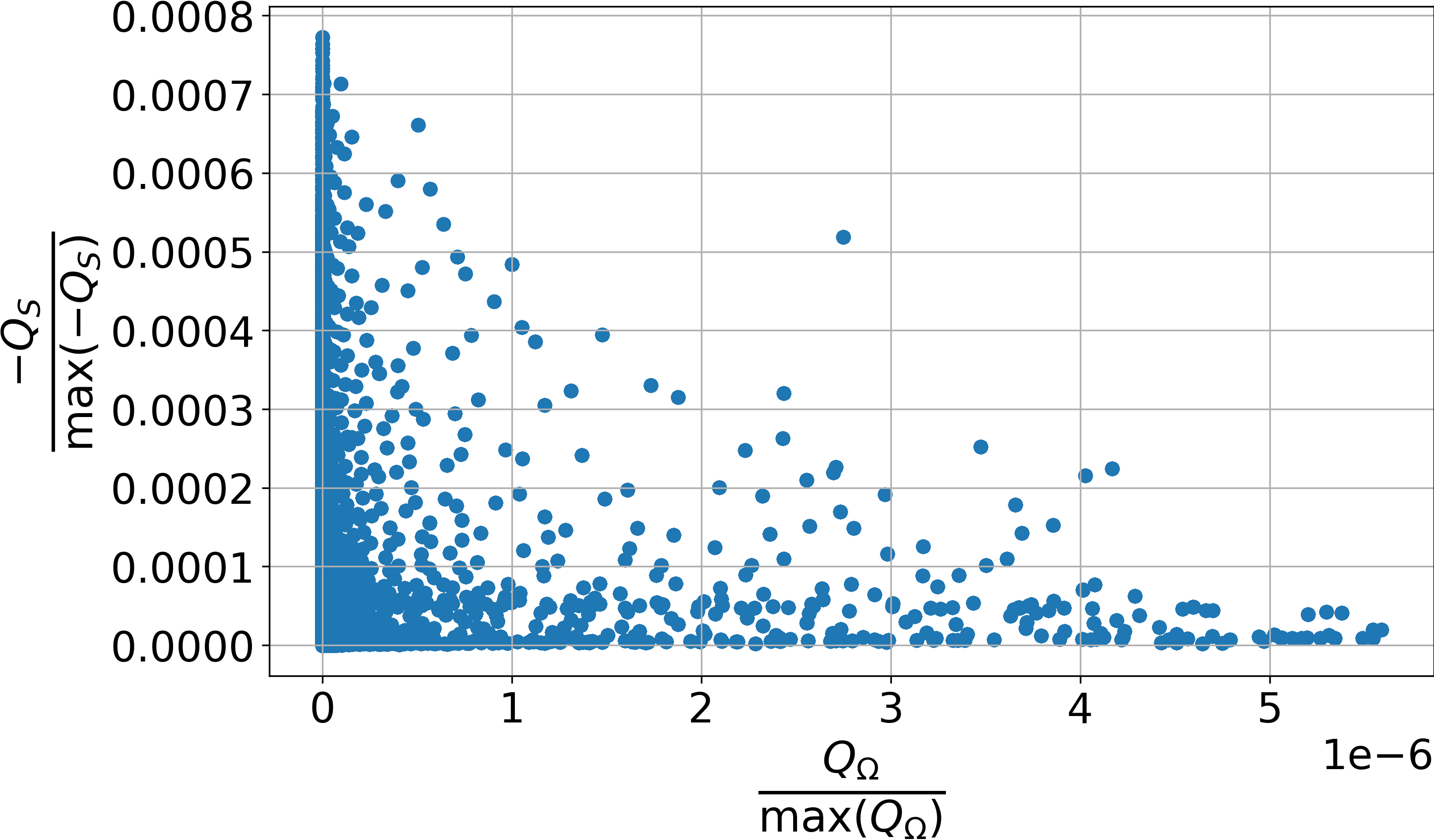}
         \caption{$(Q_\Omega,-Q_S)$ scatter plot in the detected outer flow region.}
         \label{Q_sVsQ_omega_inviscid_Re=40}
     \end{subfigure}
     \hfill
     \begin{subfigure}[h!]{0.45\textwidth}
         \centering
         \includegraphics[width=\textwidth]{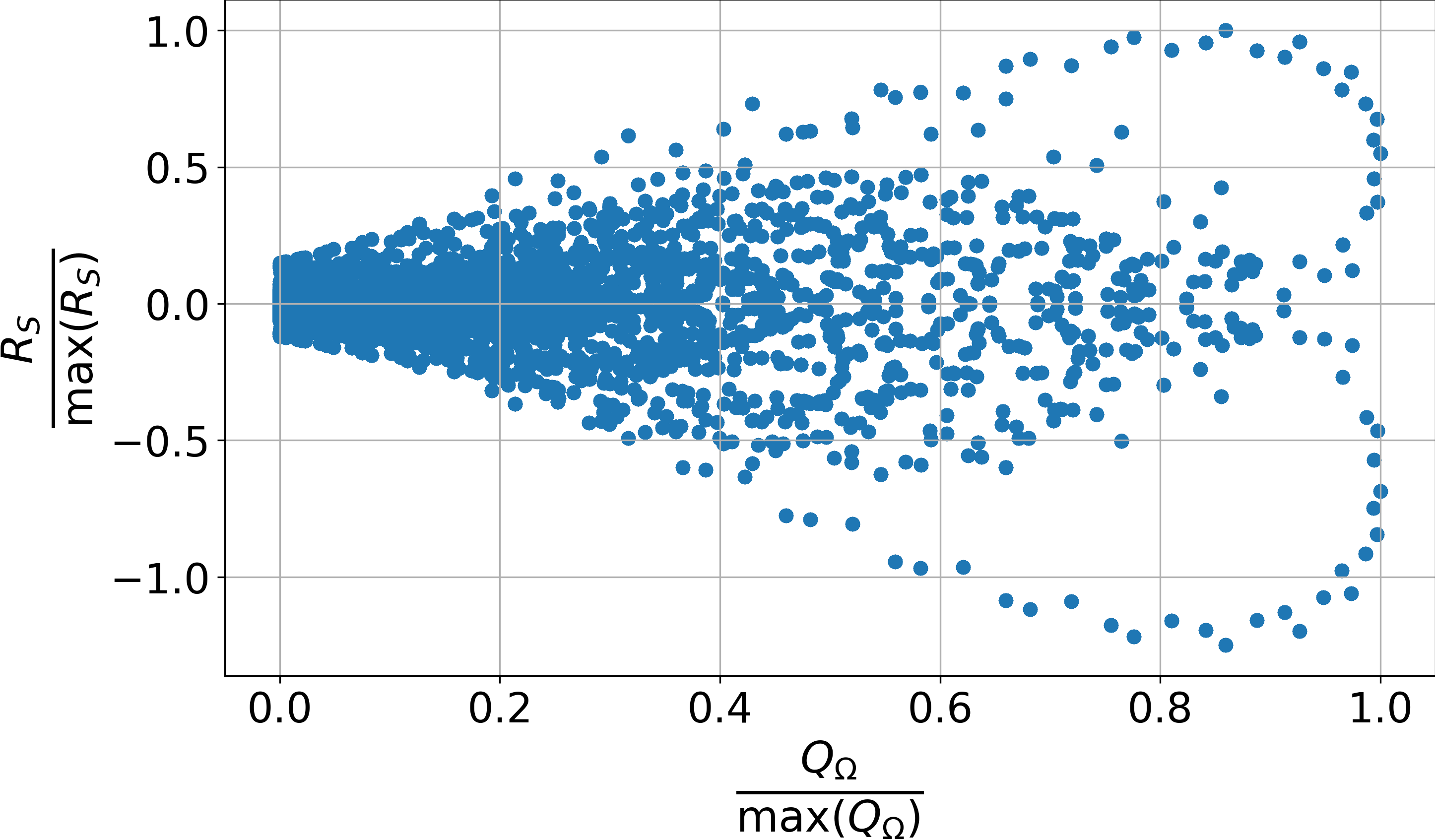}
         \caption{($Q_\Omega,R_S)$ scatter plot in the detected boundary layer and wake region.}
         \label{R_sVsQ_omega_viscous_Re=40}   
     \end{subfigure}
     \hfill
     \begin{subfigure}[h!]{0.45\textwidth}
       \centering
         \includegraphics[width=\textwidth]{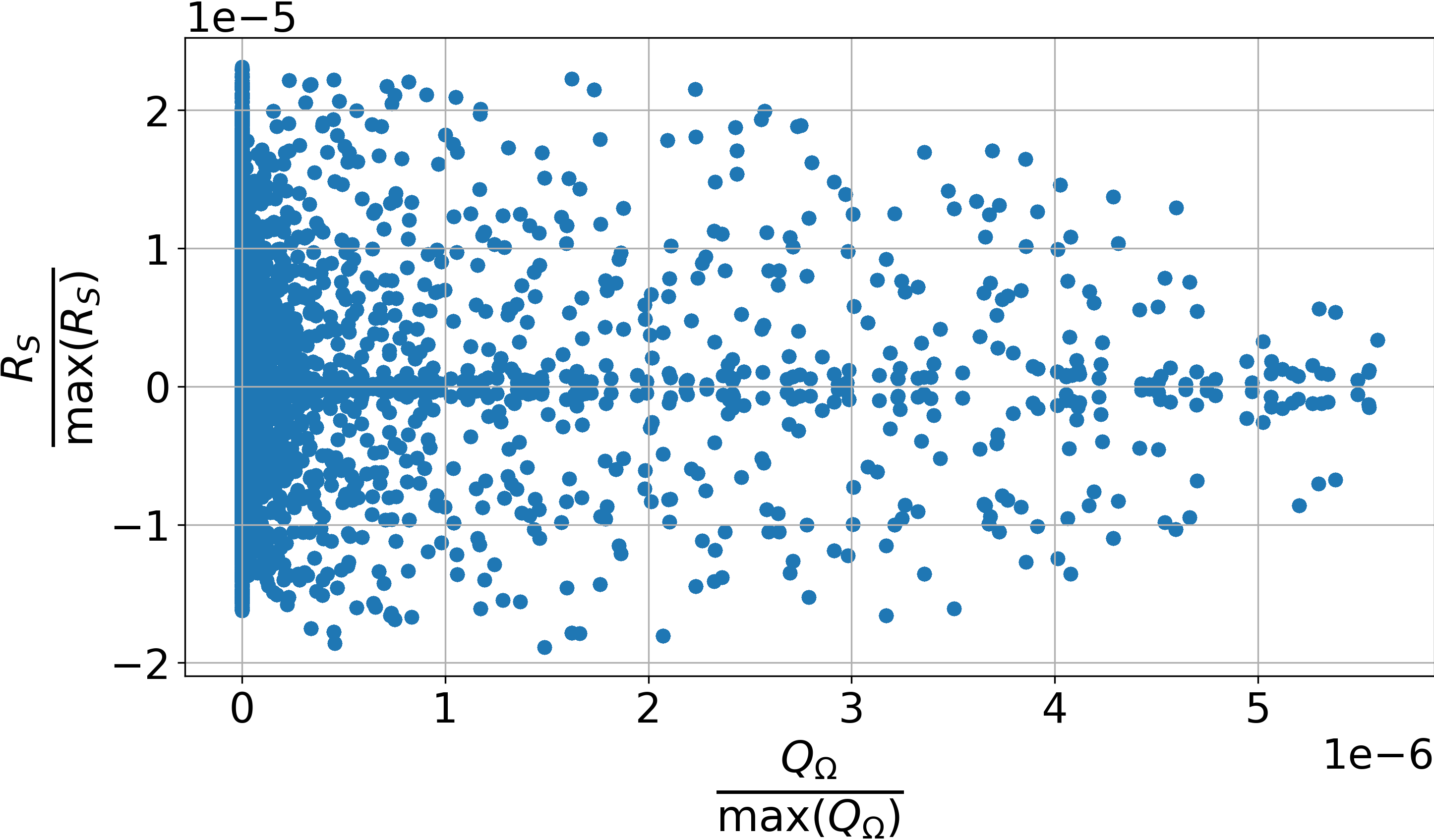}
         \caption{($Q_\Omega,R_S)$ scatter plot in the detected outer flow region.}
         \label{R_sVsQ_omega_inviscid_Re=40}   
     \end{subfigure}
     \hfill
     \begin{subfigure}{0.45\textwidth}
        \centering
         \includegraphics[width=\textwidth]{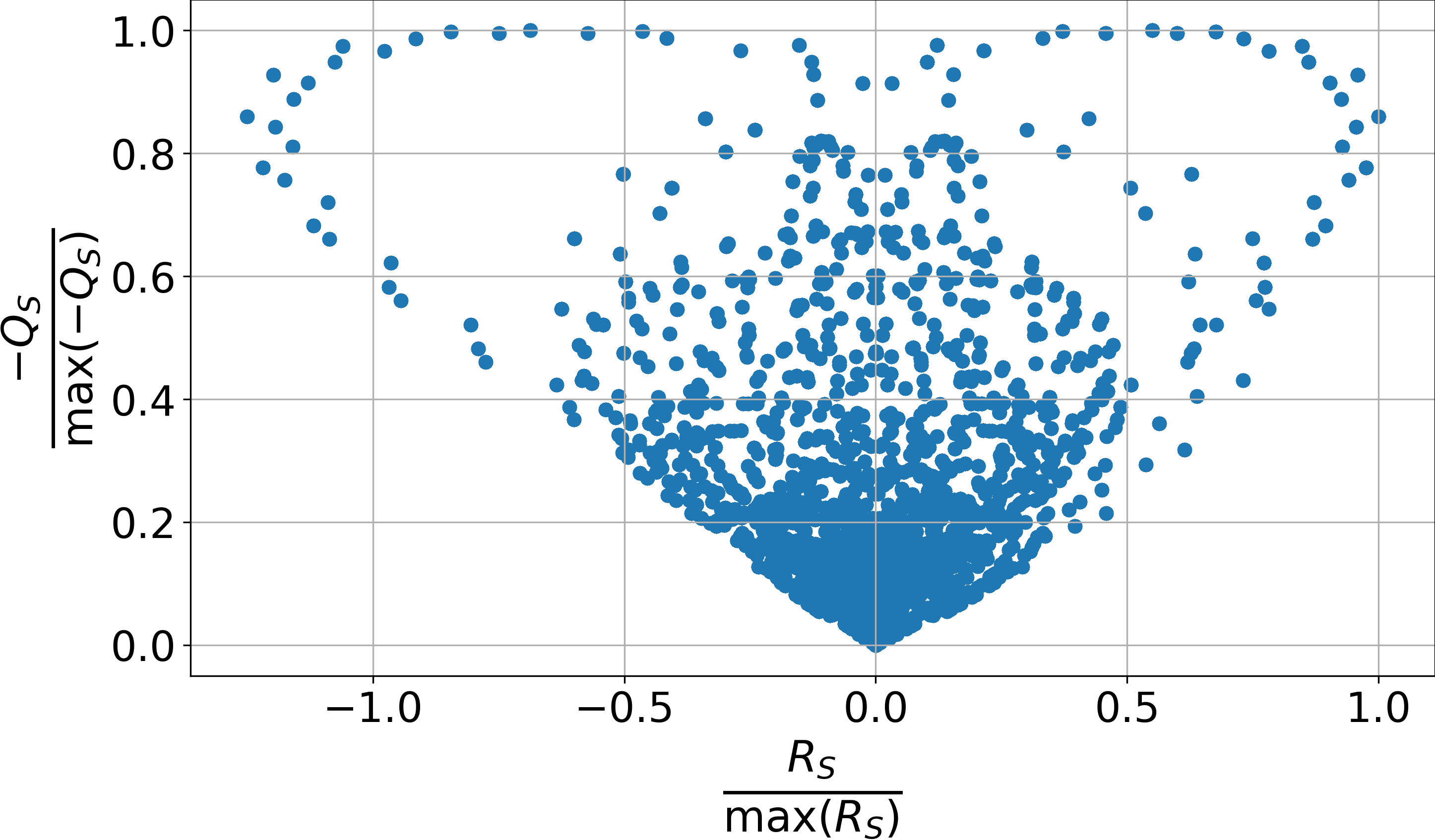}
         \caption{$(R_S,-Q_S)$ scatter plot in the detected boundary layer and wake region.}
         \label{R_sVsQ_s_viscous_Re=40}
     \end{subfigure}
     \hfill
     \begin{subfigure}{0.45\textwidth}
          \centering
         \includegraphics[width=\textwidth]{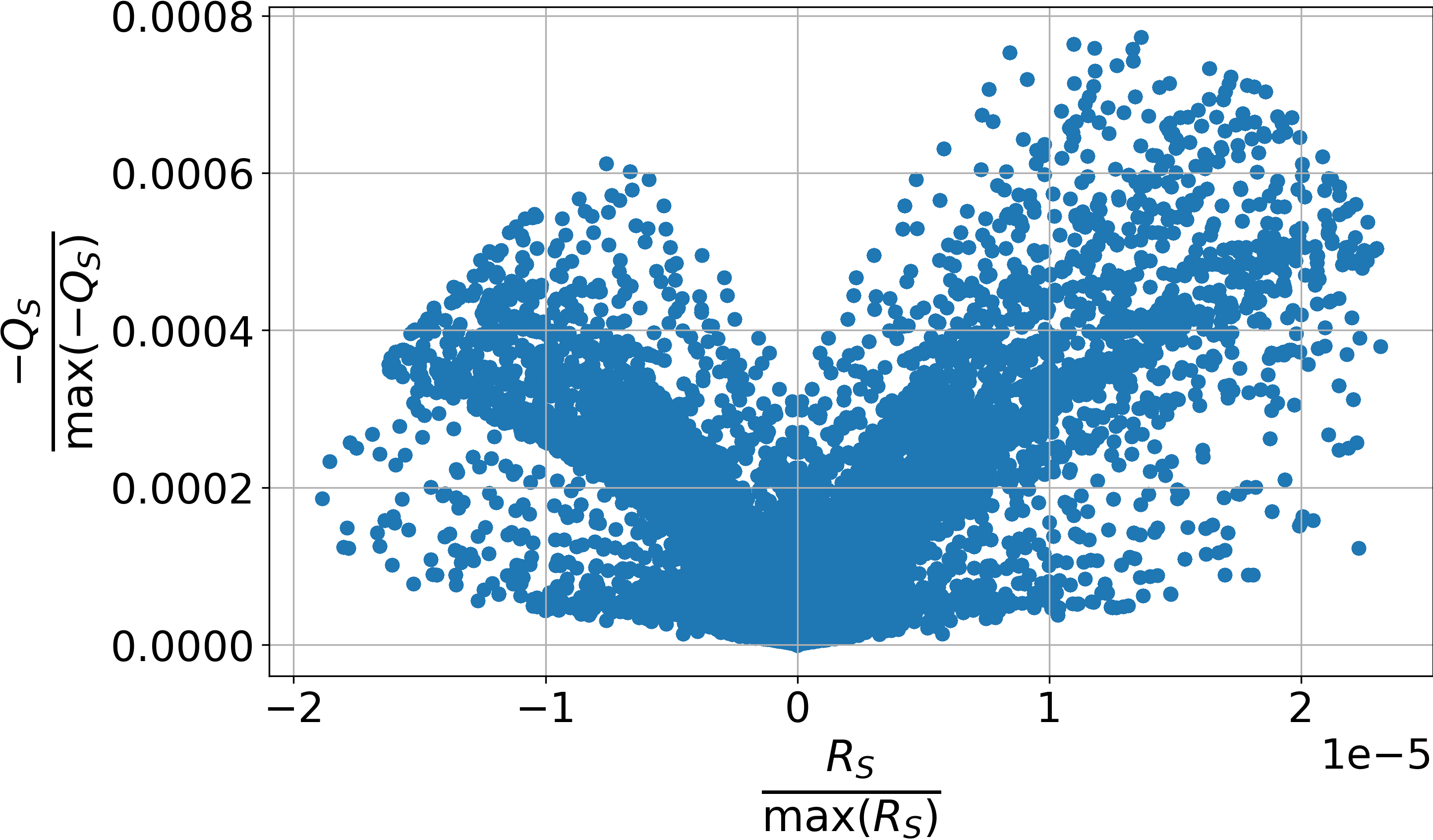}
         \caption{$(R_S,-Q_S)$ scatter plot in the detected outer flow region.}
         \label{R_sVsQ_s_inviscid_Re=40}
     \end{subfigure}
     \caption{Scatter plot of $(Q_\Omega, -Q_S)$,
     $(Q_\Omega,R_S)$ and $(R_S, -Q_S)$ in the detected regions by the GMM clustering.}
     \label{Scatter_plot_40}
\end{figure}

To further investigate the detected regions by the GMM and quantify the accuracy of the clustering, we present the scatter plots of $Q_S$, $R_S$ and $Q_\Omega$ in each clustered region. Figure~\ref{Scatter_plot_40} presents the scatter plots of the invariants of strain and rotational rate tensors maps $(Q_\Omega,-Q_S)$, $( Q_\Omega,R_S)$ and $(R_S,-Q_S)$. It is important to mention that the data presented in the plots are scaled with the maximum of the data in the entire domain. By doing so, we can easily see that if the data are close to one, the degrees of freedom with large values of the invariant are included in the selected region. However, if the degrees of freedom have low values for the scaled invariants, then the region does not contain significant viscous/turbulent effects.

The scatter plot of $ (Q_\Omega, -Q_S)$ in figure~\ref{Q_sVsQ_omega_viscous_Re=40} shows that high viscous dissipation and vorticity values are concentrated in the clustered region where the boundary layer and wake are included, while viscous dissipation and enstrophy are negligible in the region identified as the outer flow, as the values of $\frac{-Q_S}{max(-Q_S)}$ and $\frac{Q_\Omega}{max(Q_\Omega)}$ are of 
order $O\left({10^{-4}}\right)$ and $O\left({10^{-6}}\right)$. 

The clustering and scatter plots provide additional information since, as explained in \cite{da_silva_2008}, in regions where $Q_\Omega \approx -Q_S$ the flow is known to have a vortex sheet shape \cite{horiuti_2005} as shown in figure~\ref{Q_sVsQ_omega_viscous_Re=40}. The scatter plots of $(Q_\Omega, R_S)$ and $(R_S, -Q_S)$ in figures~\ref{R_sVsQ_omega_viscous_Re=40} and \ref{R_sVsQ_s_viscous_Re=40} indicate that the detected boundary layer and wake are characterised by high rates of strain production and destruction in the flow field. On the contrary, in the outer flow region no strain production or destruction occurs, as shown in figures~\ref{R_sVsQ_omega_inviscid_Re=40} and \ref{R_sVsQ_s_inviscid_Re=40}.

\subsection{Cylinder flow at Re=3900}
In the case of the flow past a circular cylinder at $Re=3900$ we use the solution field at a time instant after the flow has been fully developed. We perform the clustering using the same feature space $ E$, see section~\ref{sec:meth}, to challenge its robustness through this turbulent case. We compare our clustering method with the classic eddy viscosity sensor $F_{\mu_t}$. The eddy viscosity sensor has been tested with different parameters $K$ from the interval $K\in[1.25, 1.75]$.
\begin{figure}
    \centering
     \begin{subfigure}[t]{0.45\textwidth}
     \includegraphics[width=\textwidth]{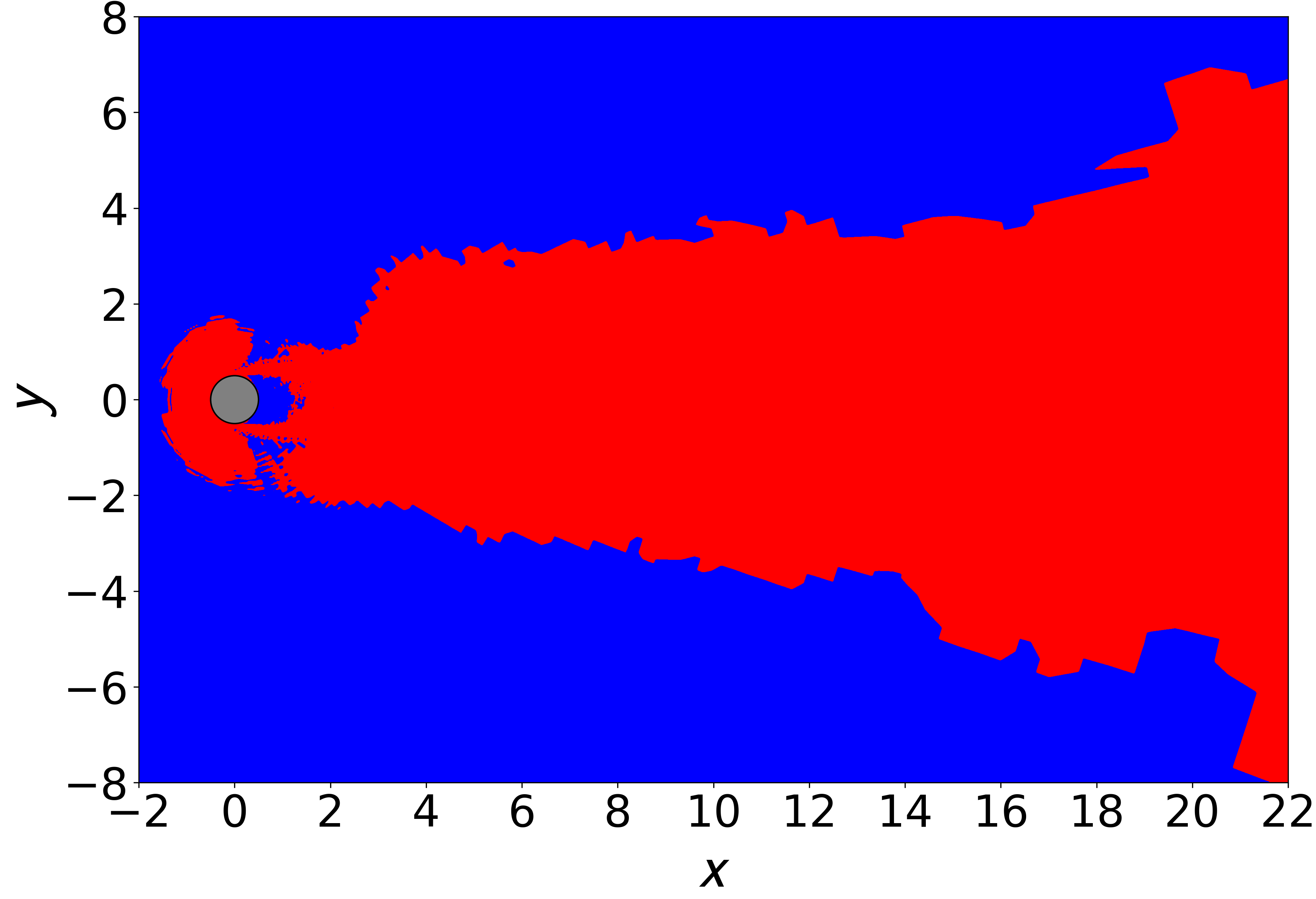}
     \caption{ }
    \label{regions_k_1_25}
    \end{subfigure}
    \hfill
      \begin{subfigure}[t]{0.45\textwidth}
     \includegraphics[width=\textwidth]{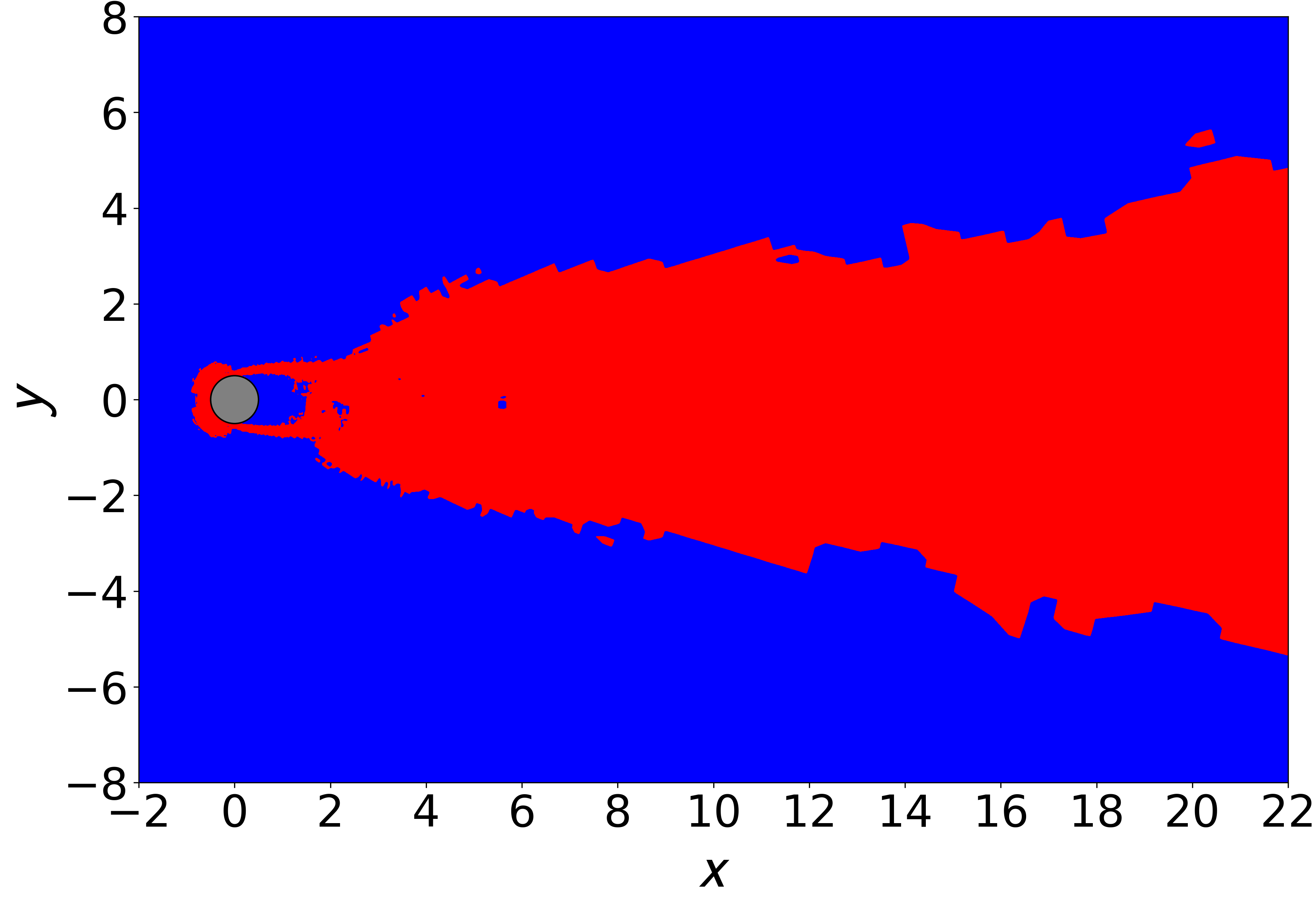}
    
    \caption{ }
    \label{regions_k_1_5}
    \end{subfigure}
    \hfill
     \begin{subfigure}[t]{0.45\textwidth}
      \includegraphics[width=\textwidth]{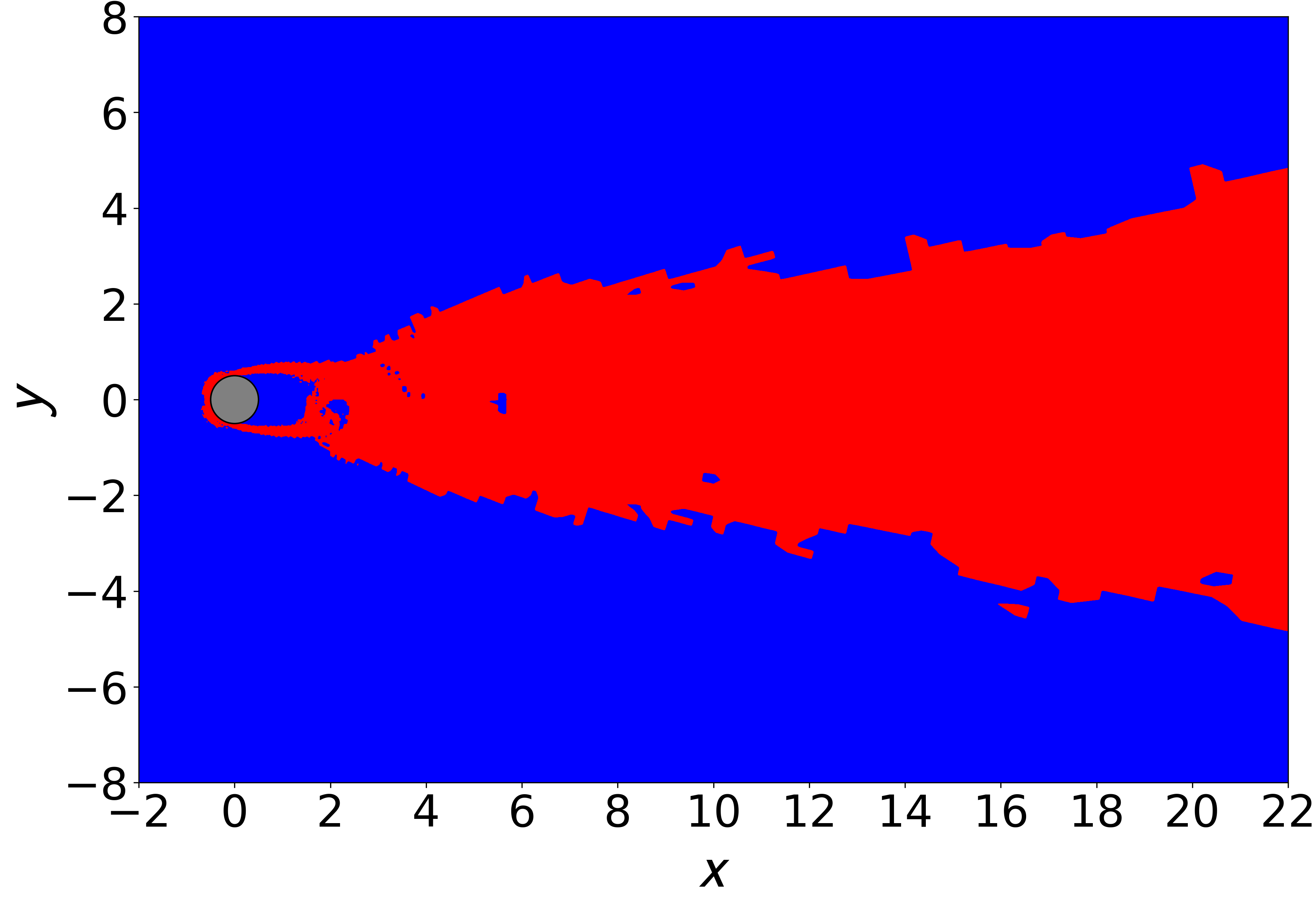}
      \caption{ }
      \label{regions_k_1_75}
     \end{subfigure}
    \hfill
    \begin{subfigure}[t]{0.45\textwidth}
     \includegraphics[width=\textwidth]{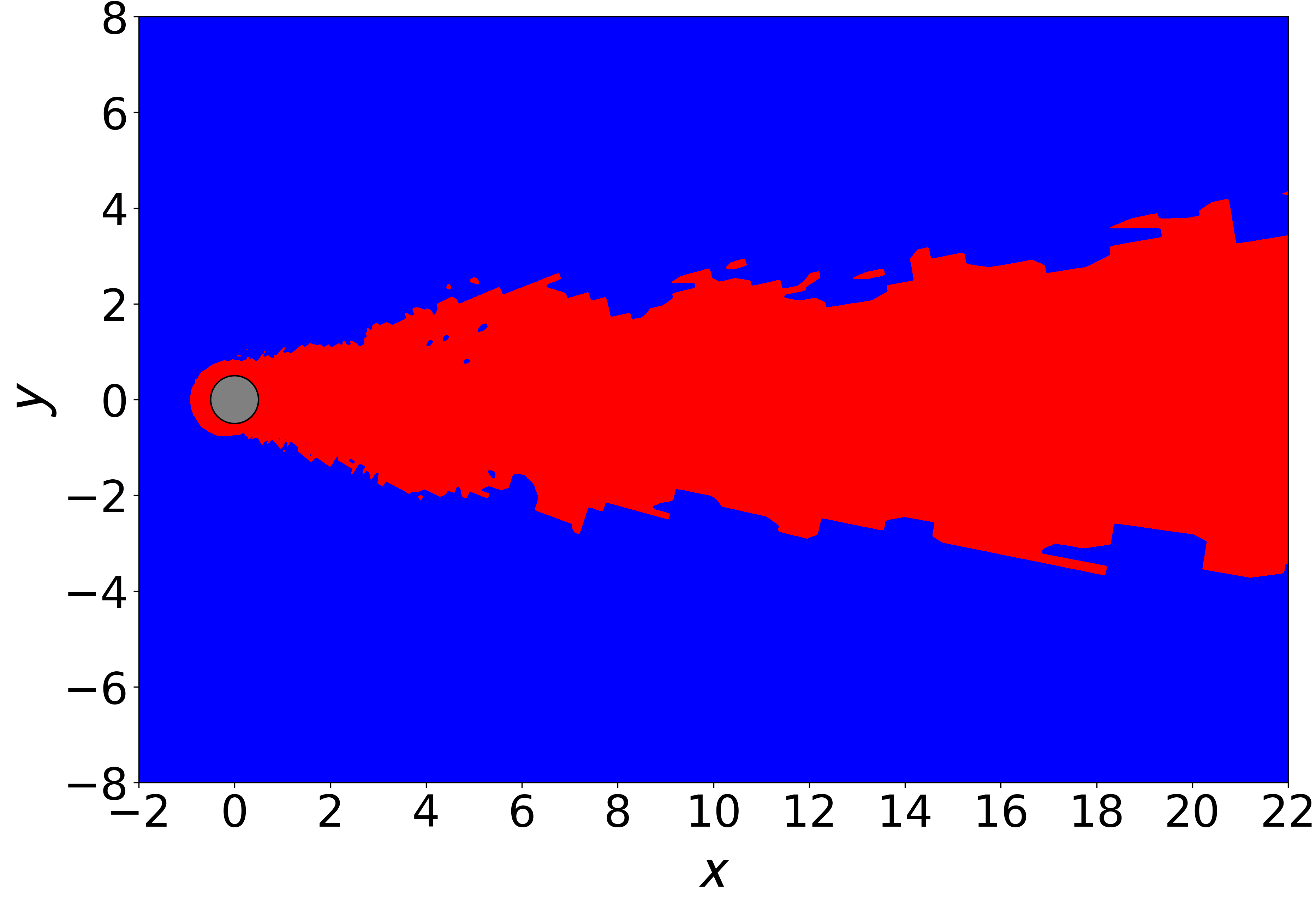}
     \caption{ }
     \label{GM_3900}
    \end{subfigure}
    \caption{Flow regions detection for flow past a cylinder at $Re=3900$ using eddy viscosity sensor $F_{\mu_t}$, $K=1.25$ (\ref{regions_k_1_25})  , $K=1.5$ (\ref{regions_k_1_5}),   $K=1.75$ (\ref{regions_k_1_75}) and GMM clustering with feature space $ E$. (\ref{GM_3900}), \textcolor{red}{Red}: Boundary layer and wake regions, \textcolor{blue}{Blue}: Outer flow region.}
    \label{regions_re_3900}
\end{figure}

The results are presented in figure~\ref{regions_k_1_75},\ref{regions_k_1_5} and \ref{regions_k_1_25} for the traditional sensors and \ref{GM_3900} for the proposed clustering. The figures show that the GMM clustering can provide similar results to the classic sensor (when tuned correctly $K\sim1.75$). Note that the classic sensor $F_{\mu_t}$ fails to identify a region in the near wake, close to the back of the cylinder, where there is no turbulent viscosity ($ \mu_t \to 0$). However, this region is dominated by viscous effects and is misidentified by $F_{\mu_t}$ as being part of the outer region. Our methodology successfully clusters this region into the viscous/turbulent regions.
The GMM clustering method, along with the feature space $E$, can overcome this issue and detects the viscous-dominated region downstream of the cylinder, as presented in figure~\ref{GM_3900}, without the need of tuning the threshold of any parameters.

\begin{figure}
     \centering
     \begin{subfigure}[h!]{0.45\textwidth}
         \centering
         \includegraphics[width=\textwidth]{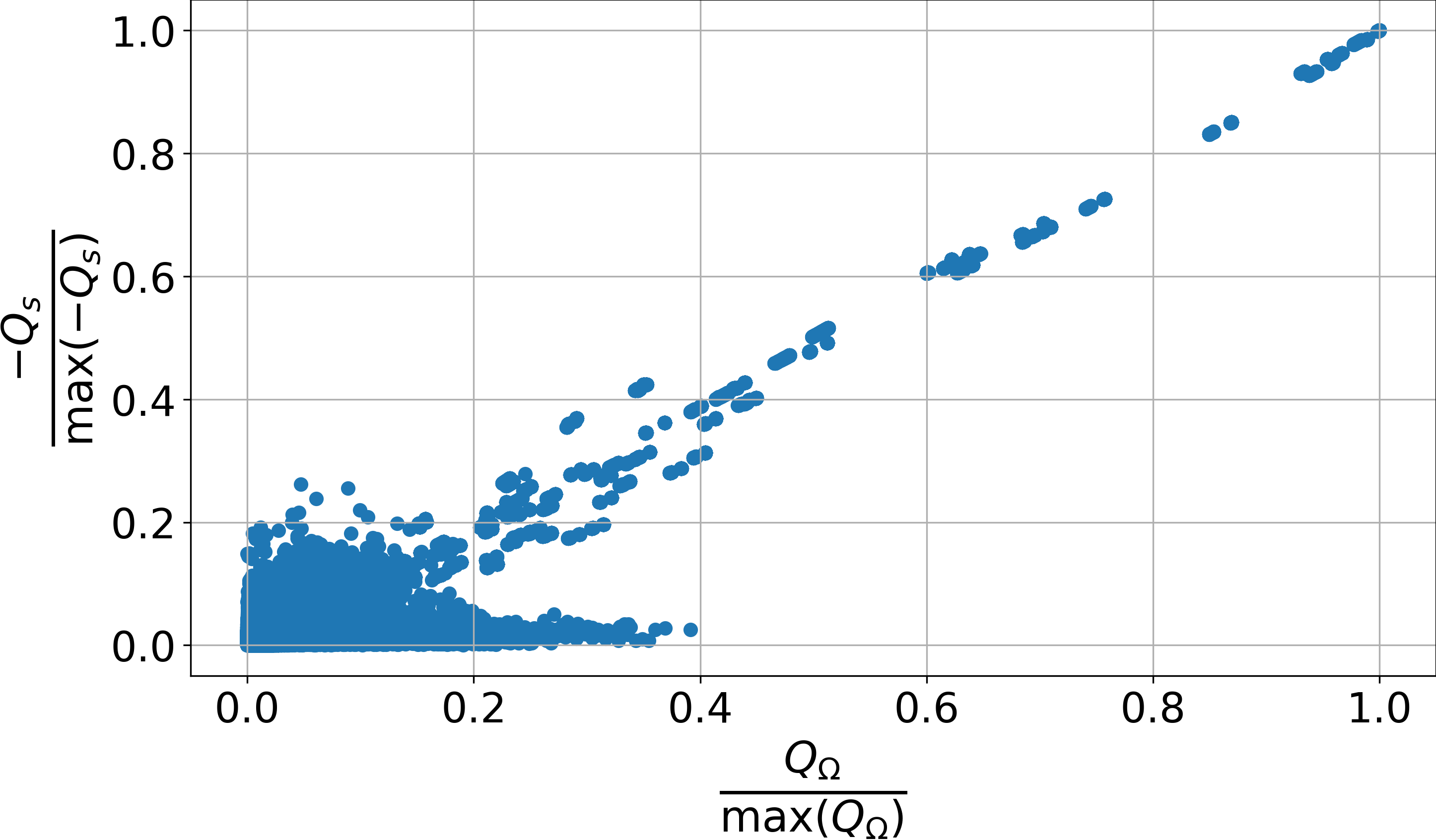}
         \caption{$(Q_\Omega,-Q_S)$ scatter plot in the detected boundary layer and wake region.}
         \label{Q_sVsQ_omega_viscous_Re=3900}
     \end{subfigure}
     \hfill
     \begin{subfigure}[h!]{0.45\textwidth}
         \centering
         \includegraphics[width=\textwidth]{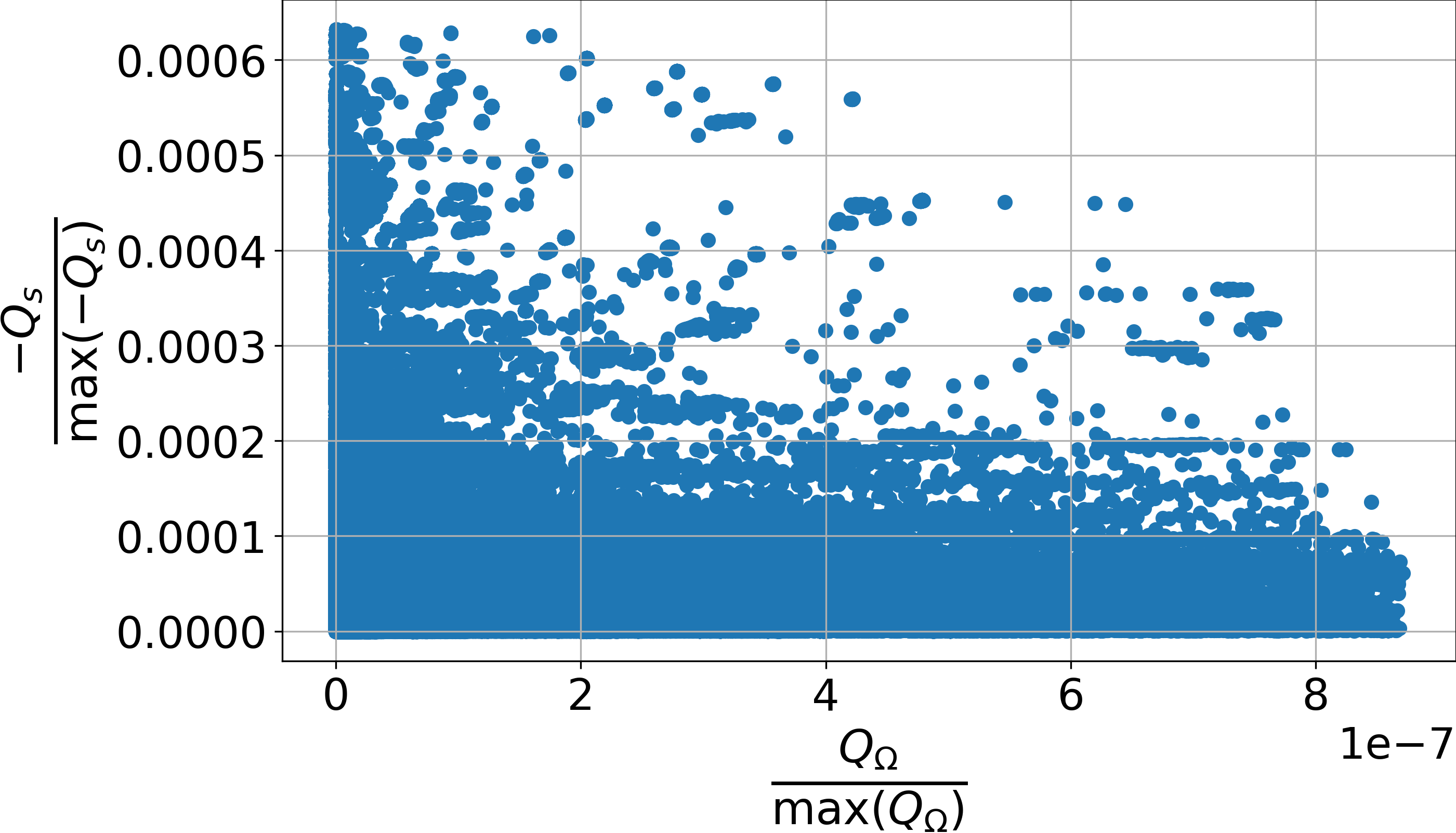}
         \caption{$(Q_\Omega,-Q_S)$ scatter plot in the detected outer flow region.}
         \label{Q_sVsQ_omega_inviscid_Re=3900}
     \end{subfigure}
     \hfill
     \begin{subfigure}[h!]{0.45\textwidth}
         \centering
         \includegraphics[width=\textwidth]{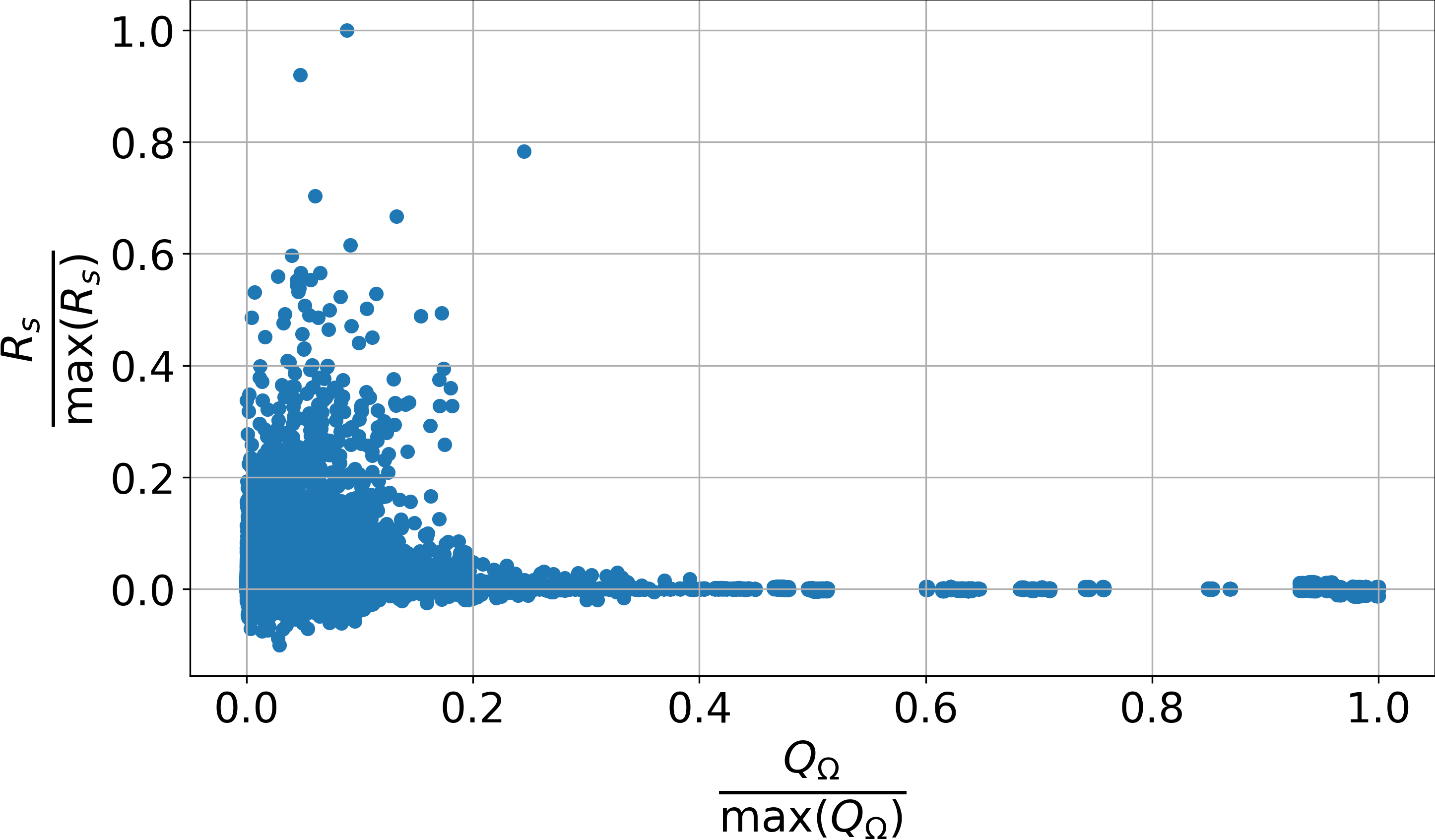}
         \caption{$(Q_\Omega, R_S)$ scatter plot in the detected boundary layer and wake region.}
         \label{R_sVsQ_omega_viscous_Re=3900}   
     \end{subfigure}
     \hfill
     \begin{subfigure}[h!]{0.45\textwidth}
       \centering
         \includegraphics[width=\textwidth]{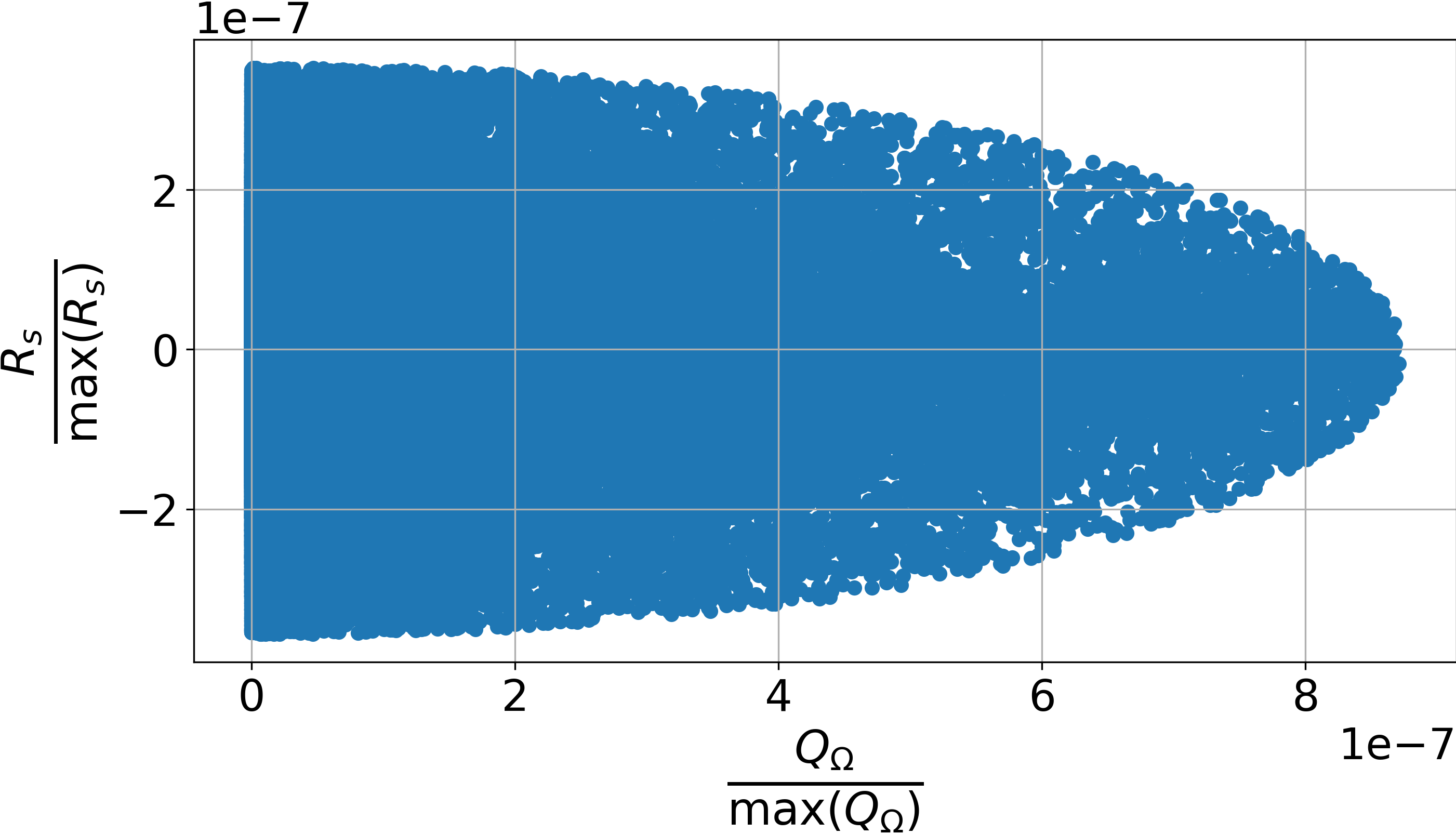}
         \caption{$(Q_\Omega, R_S)$ scatter plot in the detected outer flow region.}
         \label{R_sVsQ_omega_inviscid_Re=3900}   
     \end{subfigure}
     \hfill
     \begin{subfigure}{0.45\textwidth}
        \centering
         \includegraphics[width=\textwidth]{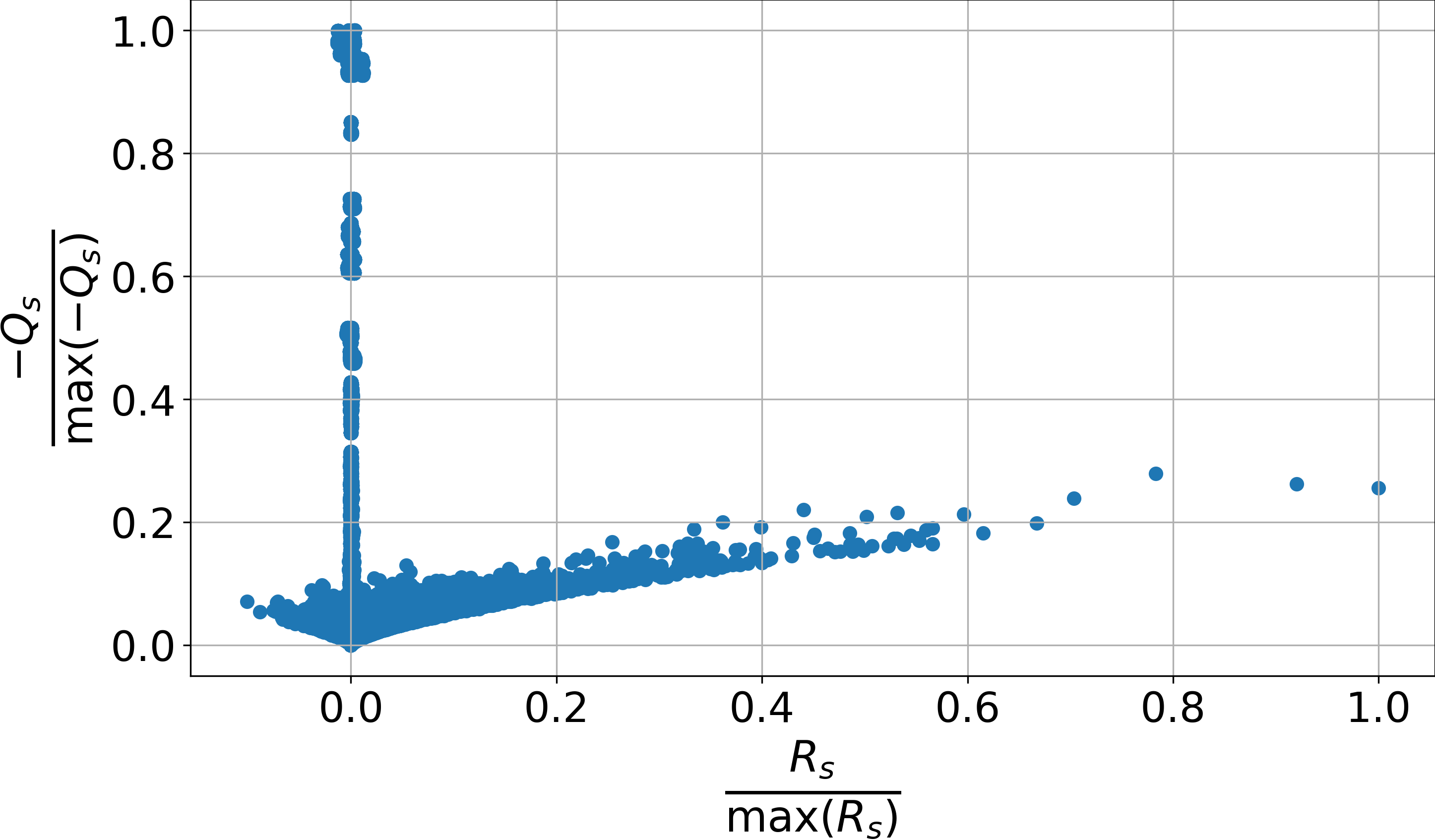}
         \caption{$(R_S, -Q_S)$ scatter plot in the detected boundary layer and wake region.}
         \label{R_sVsQ_s_viscous_Re=3900}
     \end{subfigure}
     \hfill
     \begin{subfigure}{0.45\textwidth}
          \centering
         \includegraphics[width=\textwidth]{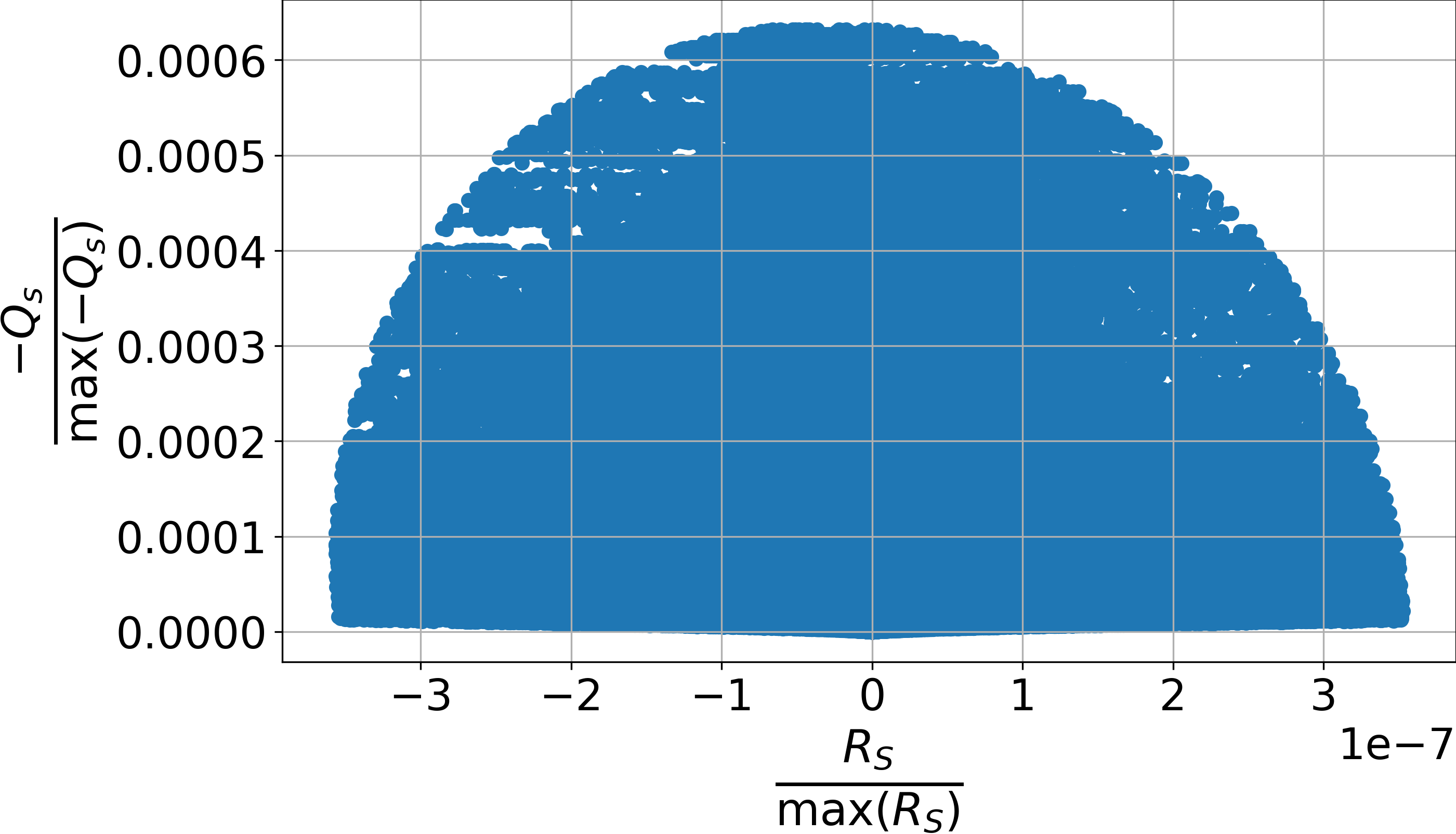}
         \caption{$(R_S, -Q_S)$ scatter plot in the detected outer flow region.}
         \label{R_sVsQ_s_inviscid_Re=3900}
     \end{subfigure}
     \caption{Scatter plot of $(Q_\Omega,R_S)$,
     $(Q_\Omega,R_S)$ and $( R_S,-Q_S)$ in the detected regions by the GMM clustering.}
     \label{Scatter_plot_3900}
\end{figure}

As in the case of the laminar flow, we analyse the scatter plots of the invariants of the strain and rotational rate tensor maps $( Q_\Omega, -Q_S)$, $( Q_\Omega, R_S)$ and $(R_S, -Q_S)$ which are presented in figure~\ref{Scatter_plot_3900}. Again, the plots are scaled with the maximum values of $Q_\Omega$,$-Q_S$ and $R_S$ in the entire flow.

The high values of viscous dissipation and enstrophy are concentrated in the detected boundary layer and wake regions, as shown in figure~\ref{Q_sVsQ_omega_viscous_Re=3900}. In the detected outer flow region, viscous dissipation and enstrophy are negligible, as the values of $\frac{-Q_S}{max(-Q_S)}$ and $\frac{Q_\Omega}{max(Q_\Omega)}$ are of 
order $O\left({10^{-4}}\right)$ and $
O\left({10^{-7}}\right)$, respectively. In addition, the clustering and scatter plots provide physical insights. Two flow structures can be observed in this case within the detected boundary layer and wake region. Following the results presented in figure~\ref{Q_sVsQ_omega_viscous_Re=3900}, the first region is where $Q_\Omega=-Q_S$ and is related to the existence of a vortex sheet shape \cite{horiuti_2005}. In the second region, enstrophy is dominant, with high values of $Q_\Omega$ and negligible values of $Q_S$, which can be interpreted as a vortex tube shape \cite{da_silva_2008}.

Figures~\ref{R_sVsQ_omega_viscous_Re=3900}~and~\ref{R_sVsQ_s_viscous_Re=3900} show that considerable strain production and destruction occur in the detected boundary layer and wake regions. In the detected outer flow region, no clear shapes/features appear as shown in figures~\ref{R_sVsQ_omega_inviscid_Re=3900} and \ref{R_sVsQ_s_inviscid_Re=3900}.

\begin{figure}
     \centering
      \begin{subfigure}[b!]{0.455\textwidth}
         \centering
\includegraphics[width=\textwidth]{Results/Re_3900/Gaussian_mixture/Q_sVsQ_omega_Inviscid_Re_3900.png}
         \caption{$(Q_\Omega,-Q_S)$ scatter plot in the detected outer flow region at $Re=3900$ with feature space $E$.}
         \label{Q_sVsQ_omega regions_re_40_outer}
     \end{subfigure}
     \hfill
     \begin{subfigure}[b!]{0.455\textwidth}
         \centering
         \includegraphics[width=\textwidth]{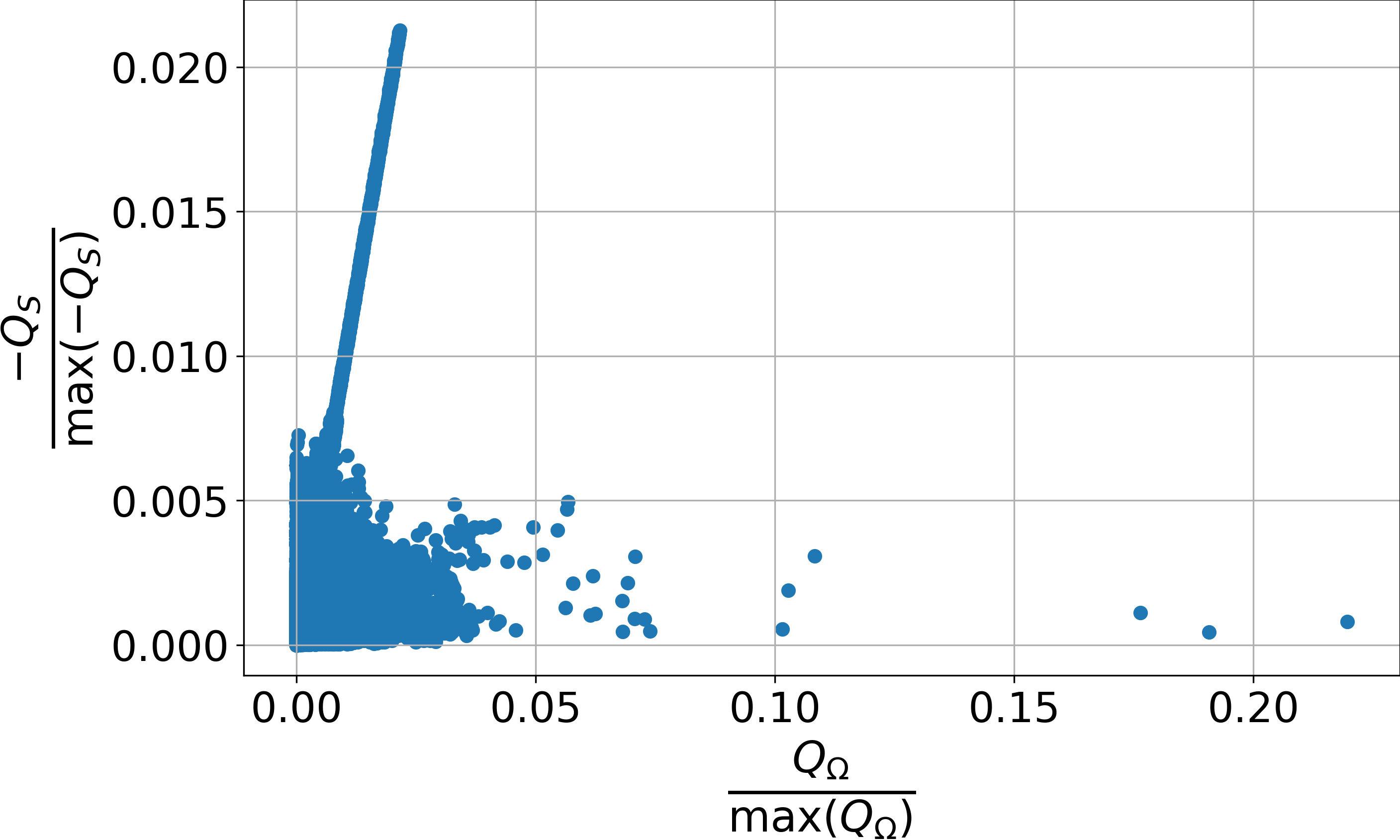}
         \caption{$(Q_\Omega,-Q_S)$ scatter plot in the detected outer flow region at $Re=3900$ with $F_{\mu_t}$. $K=1.75$.}
         \label{Q_sVsQ_omega_regions_re=40_mu_t_1.75}
     \end{subfigure}
    \caption{$(Q_\Omega,-Q_S)$ scatter plot in the detected outer flow region at $Re=3900$ with $F_{\mu_t}$.  $K=1.75$ and GMM clustering with feature space $E$.}
    \label{GMMvsSensor_3900}
     \end{figure}

We also provide scatter plots for the traditional sensors to show their inadequacy in clustering regions. Figure~\ref{GMMvsSensor_3900} presents the scatter plot of $( Q_\Omega, -Q_S)$ in the outer region detected using the eddy viscosity sensor $F_{\mu_t}$ and the GMM clustering with feature space $E$. The threshold parameter $K$ of the turbulent viscosity sensor $F_{\mu_t}$ is set to $K=1.75$. The comparison of the scatter plots of the $( Q_\Omega, -Q_S)$ map shows that the GMM clustering outperforms the traditional sensor.

As shown in figures~\ref{regions_k_1_25}, \ref{regions_k_1_5} and \ref{regions_k_1_75}, $F_{\mu_t}$ is unable to cluster regions where $\mu_t \to 0$ downstream of the cylinder into the viscous dominated region. This region is characterised by non-negligible viscous dissipation. The scatter plot \ref{Q_sVsQ_omega_regions_re=40_mu_t_1.75} shows that a relatively high viscous dissipation is present in the region classified as outer flow by $F_{\mu_t}$ as $\frac{-Q_S}{max(-Q_S)}$ reaches values of 0.02. This corresponds to the misidentified regions downstream of the cylinder in figures~\ref{regions_k_1_25}, \ref{regions_k_1_5} and \ref{regions_k_1_75}. For GMM clustering with the feature space $E$, $\frac{-Q_S}{max(-Q_S)}$ is of the order $O\left({10^{-4}}\right)$ in the detected outer region. High enstrophy values can also be observed in the outer region detected using $F_{\mu_t}$ as $\frac{Q_\Omega}{max(Q_\Omega)}$ reaches $0.2$ compared to the enstrophy values in the outer region detected with GMM clustering where this value is of order $O\left({10^{-7}}\right)$, see figure~\ref{Q_sVsQ_omega regions_re_40_outer}.


\subsection{A note on the relevance of the invariants}\label{sec:compress}
In this section we discuss the relevance of using each of the invariants in the feature space $ E$. In addition, we will verify that the selected feature space includes the minimum set of variables necessary to detect the boundary layer and wake region, and that when this space is reduced (to only two invariants) we cannot capture the correct flow regions. Each one of the invariants in the feature space $E$ provide specific information about the physical mechanisms and all are deemed to be necessary. To check this hypothesis, we train the GMM clustering with $(Q_S,R_S)$ or $(Q_S, Q_\Omega)$ or 
$(R_S, Q_\Omega)$ for both Reynolds numbers (laminar and turbulent cases, see section \ref{sec:simulation}). The new (reduced space) regions will be compared with the ones obtained when using the original feature space $ E$.

\subsubsection{Feature space  $(\boldsymbol{Q_S},\boldsymbol{R_S})$}
\begin{figure}
     \centering
     \begin{subfigure}[H]{0.37\textwidth}
         \centering
         \includegraphics[width=\textwidth]{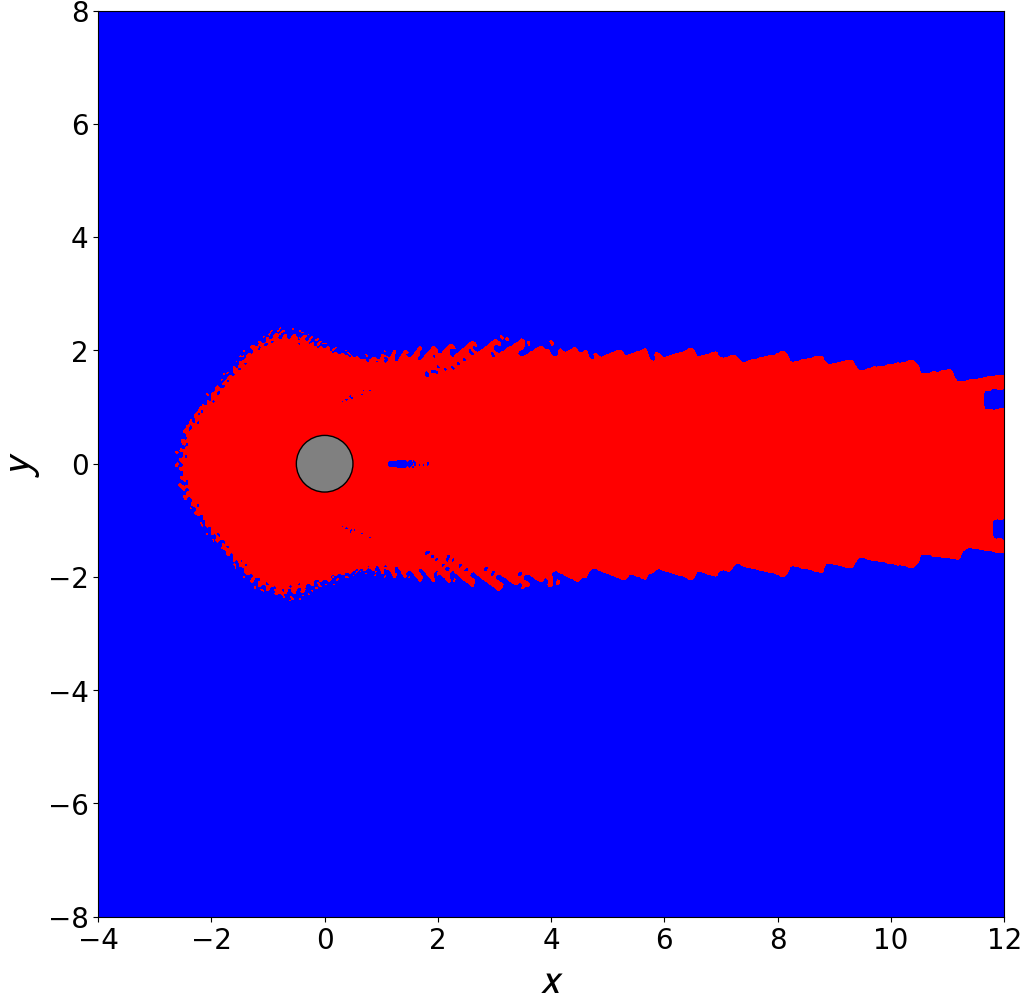}
         \caption{ }
         \label{Q_sVsR_s regions_re_40}
     \end{subfigure}
     \hfill
     \begin{subfigure}[H]{0.53\textwidth}
         \centering
         \includegraphics[width=\textwidth]{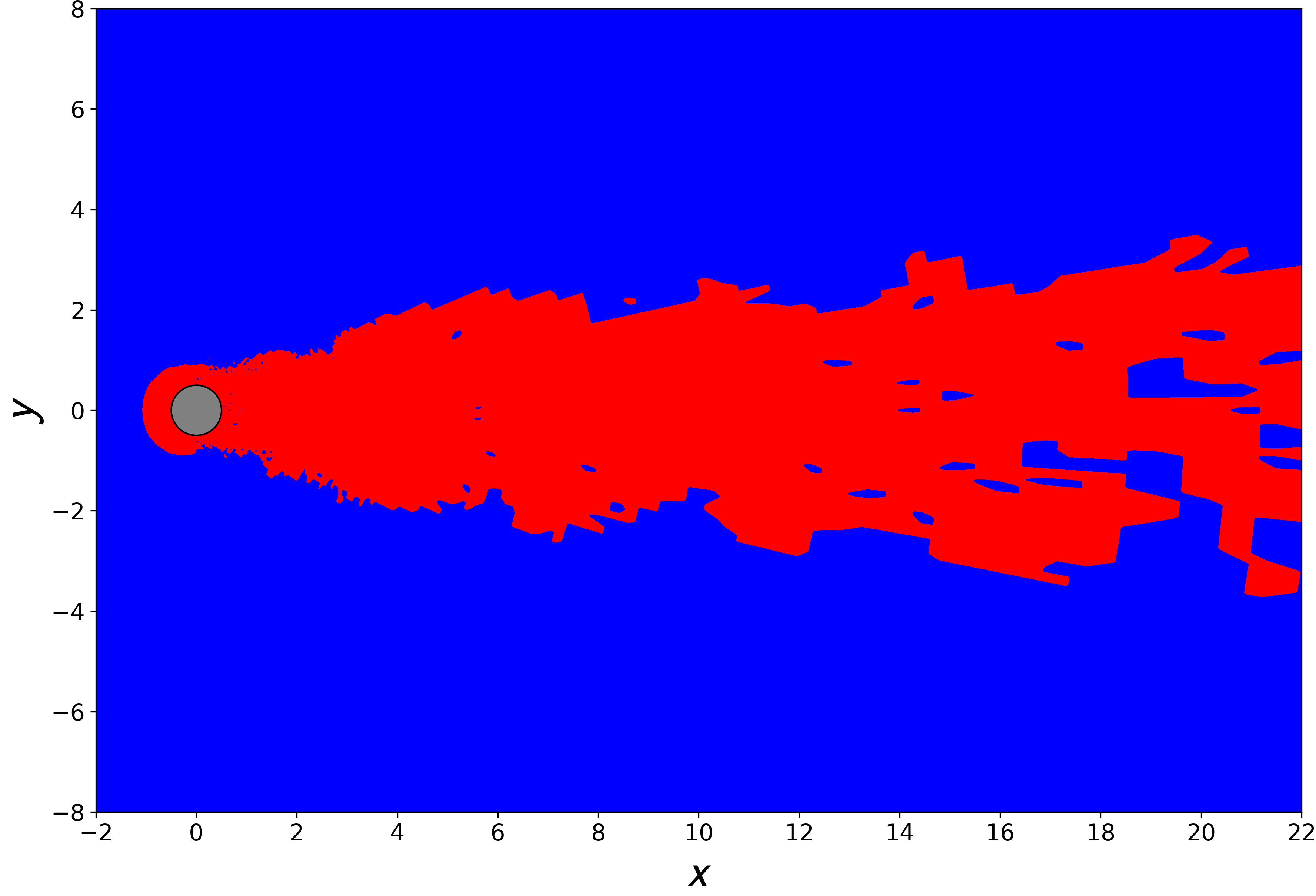}
         \caption{ }
         \label{R_sVsQ_s_regions_re=3900}
     \end{subfigure}
     \caption{Flow regions detected by GMM for flow past a cylinder at $Re=40$. (\ref{Q_sVsR_s regions_re_40}) and $Re=3900$. (\ref{R_sVsQ_s_regions_re=3900}) using  $(Q_S, R_S)$, \textcolor{red}{Red}: Boundary layer and wake regions, \textcolor{blue}{Blue}: Outer flow region.}
    \label{regions_Q_s_R_s}
 \end{figure}

The results in figures~\ref{Q_sVsQ_omega_viscous_Re=40} ($Re=40$) and \ref{Q_sVsQ_omega_viscous_Re=3900} ($Re=3900$) show that there is a strong correlation between $Q_S$ and $Q_\Omega$ in the detected boundary layer and wake region due to the presence of vortex sheet structures. Performing the clustering using $(Q_S, R_S)$, without considering the effect of $Q_\Omega$ in the feature space, we retrieve the results illustrated in figure~\ref{regions_Q_s_R_s}. 

\begin{figure}
     \centering
      \begin{subfigure}[H]{0.455\textwidth}
         \centering
\includegraphics[width=\textwidth]{Results/Q_omegaVsQ_s_Inviscid_Region_Re_40.png}
         \caption{$(Q_\Omega, -Q_S)$ scatter plot in the detected outer flow region at $Re=40$ with feature space $E$.}
         \label{Q_sVsR_s regions_re_40_E}
     \end{subfigure}
     \hfill
     \begin{subfigure}[H]{0.455\textwidth}
         \centering
         \includegraphics[width=\textwidth]{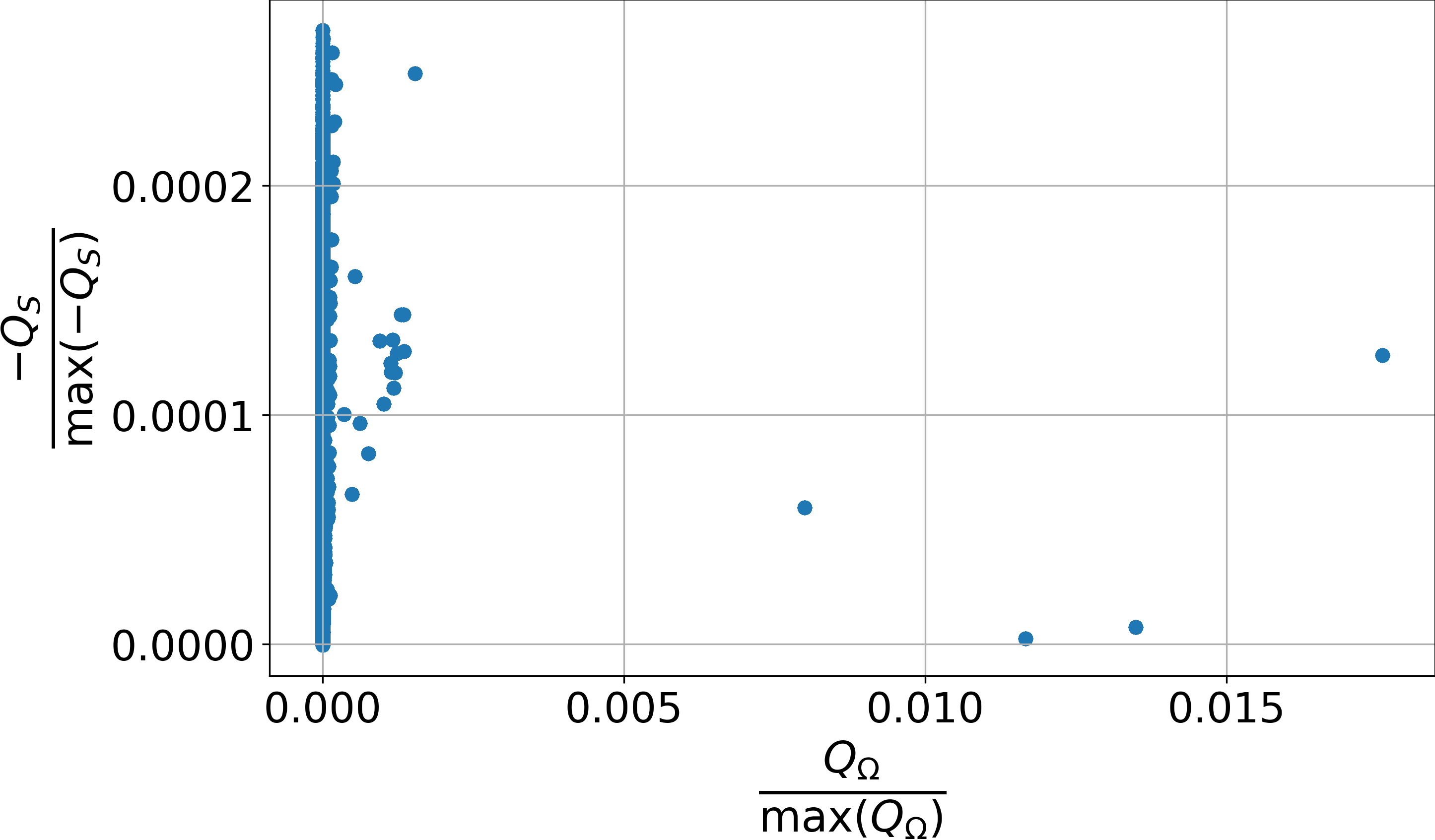}
         \caption{$(Q_\Omega, -Q_S)$ scatter plot in the detected outer flow region at $Re=40$ with $(Q_S, R_S)$.}
         \label{Q_sVsQ_omega_regions_re=40_R_s_Q_s}
     \end{subfigure}
     \begin{subfigure}[h!]{0.455\textwidth}
         \centering
         \includegraphics[width=\textwidth]{Results/Re_3900/Gaussian_mixture/Q_sVsQ_omega_Inviscid_Re_3900.png}
         \caption{$(Q_\Omega,-Q_S)$ scatter plot in the detected outer flow region at $Re=3900$ with feature space $E$.}
         \label{Q_sVsR_s regions_re_3900_E}
     \end{subfigure}
     \hfill
     \begin{subfigure}[h!]{0.455\textwidth}
         \centering
         \includegraphics[width=\textwidth]{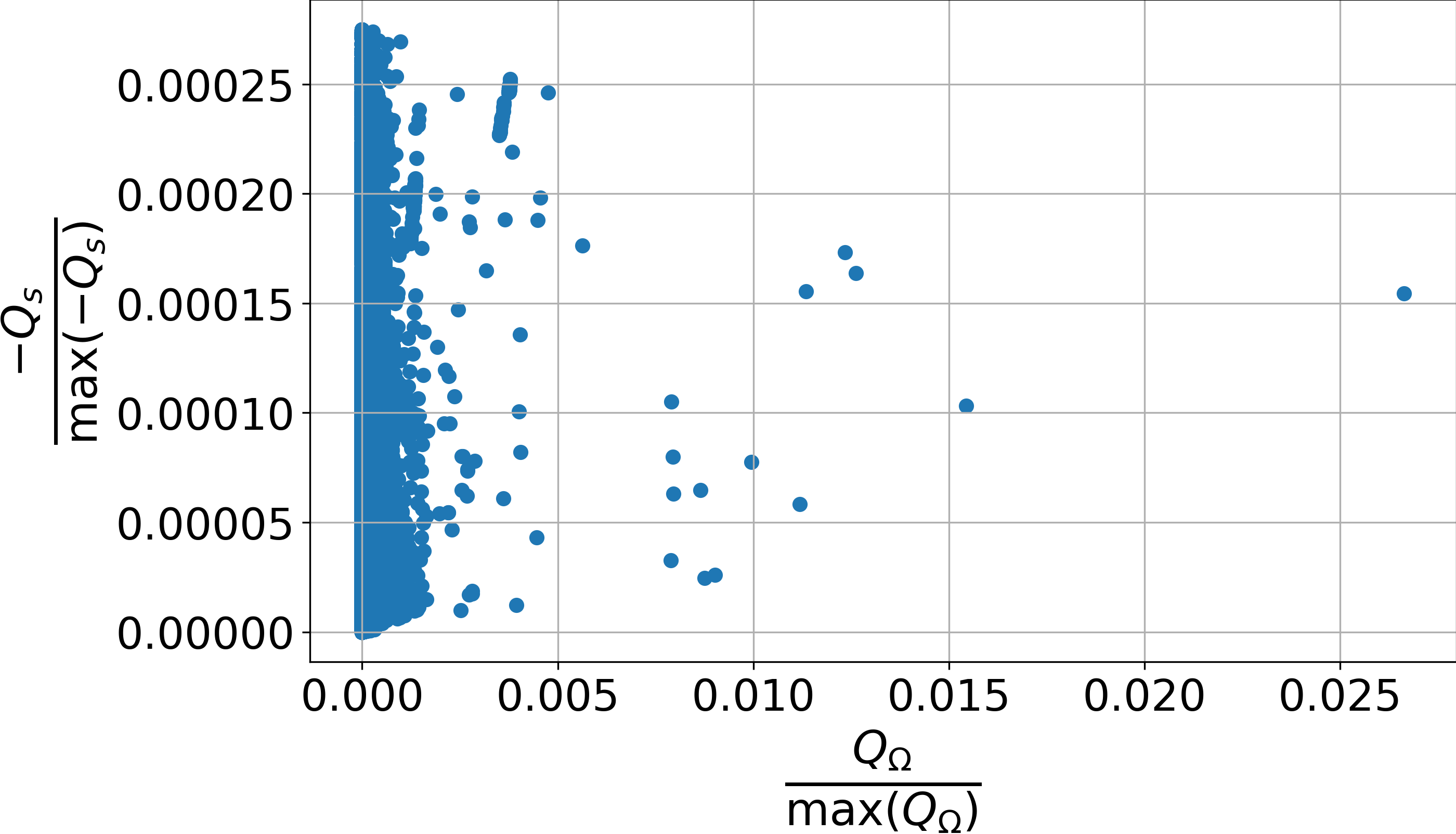}
         \caption{$(Q_\Omega, -Q_S)$ scatter plot in the detected outer flow region at $Re=3900$ with $(Q_S, R_S)$.}
         \label{Q_sVsQ_omega_regions_re=3900_R_s_Q_s}
     \end{subfigure}
     \caption{($Q_\Omega, -Q_S$) scatter plot in the detected outer flow region by GMM using the feature space $E$ and  ($Q_S, R_S$).}
     \label{scatter_plot_Q_s_R_s_regions}
     \end{figure}

The results in  figure~\ref{regions_Q_s_R_s} show patches within the wake region that are classified as outer flow region (namely in the turbulent case). This is due to the fact that within the wake, enstrophy dominated regions cannot be identified when $Q_\Omega$ is excluded from the feature space used to train the GMM. In figure~\ref{scatter_plot_Q_s_R_s_regions}, we compare the enstrophy values for the outer flow region when detecting it using $(Q_S, R_S)$ and when the clustering was performed with the feature space $E$ for the two test cases under consideration.

In figures~\ref{Q_sVsQ_omega_regions_re=40_R_s_Q_s} and~\ref{Q_sVsQ_omega_regions_re=3900_R_s_Q_s} the value of $\frac{Q_\Omega}{max(Q_\Omega)}$ reaches $0.015$ for $Re=40$ and $0.025$ for $Re=3900$ when using $(Q_S,R_S)$ for detection. On the contrary, using the feature space $E$ leads to enstrophy values which are of the order $O\left({10^{-6}}\right)$.

\subsubsection{Feature space $(\boldsymbol{Q_S},\boldsymbol{Q_\Omega})$}
\begin{figure}
     \centering
     \begin{subfigure}[b!]{0.37\textwidth}
         \centering
         \includegraphics[width=\textwidth]{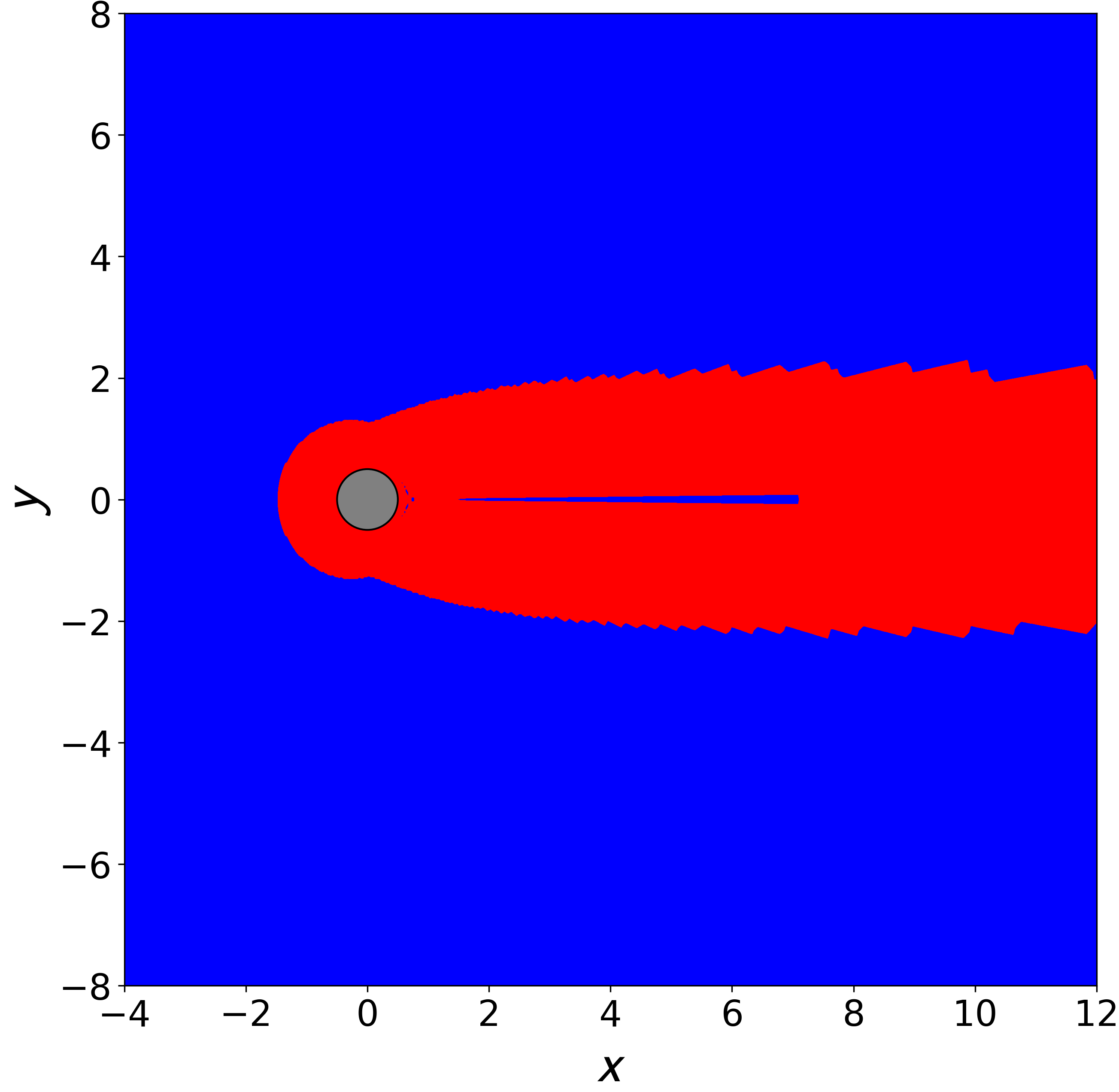}
         \caption{ }
         \label{Q_sVsQ_omega regions_re_40}
     \end{subfigure}
     \hfill
     \begin{subfigure}[b!]{0.53\textwidth}
         \centering
         \includegraphics[width=\textwidth]{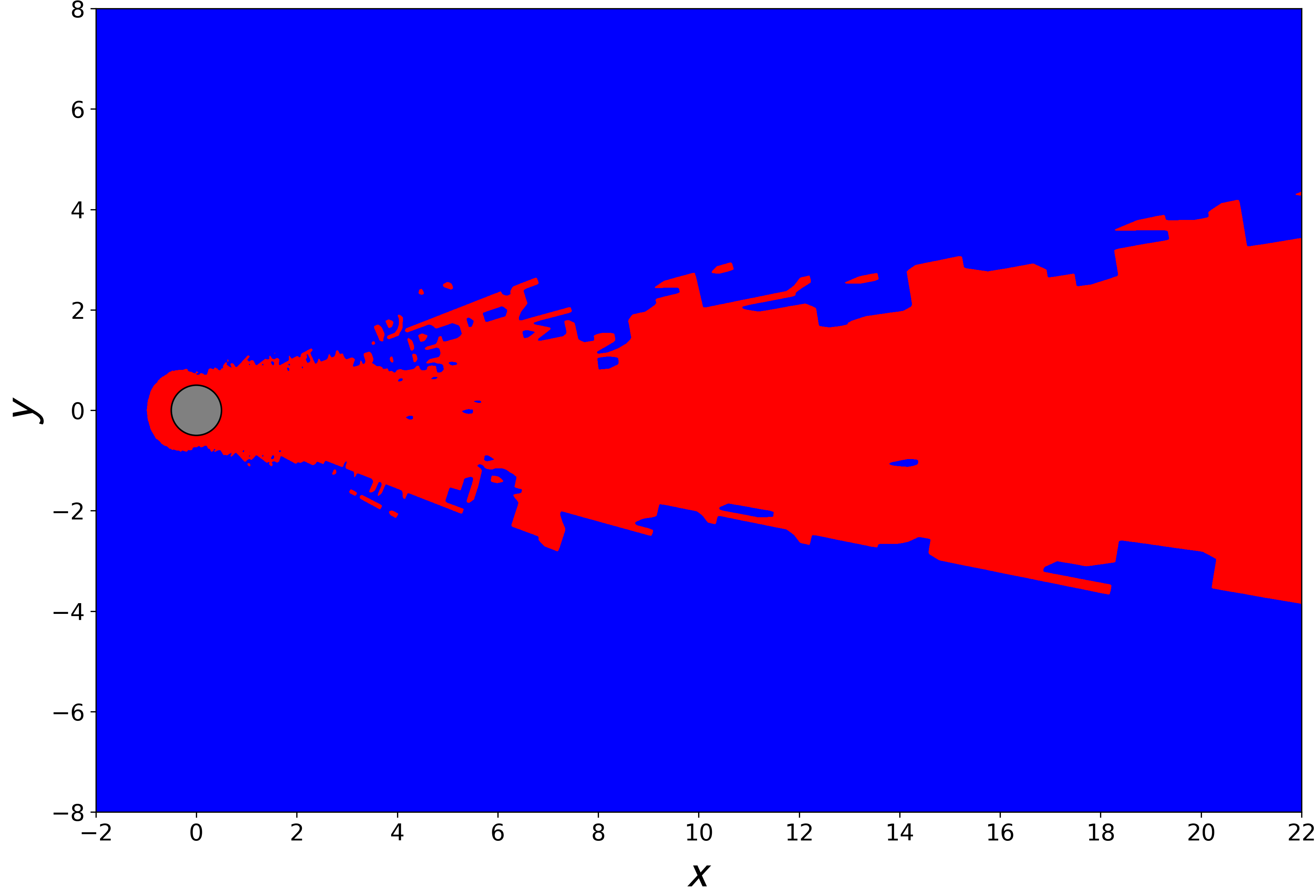}
         \caption{ }
         \label{Q_sVsQ_omega_regions_re=3900}
     \end{subfigure}
     \caption{Flow regions detected by GMM for flow past a cylinder at $Re=40$. (\ref{Q_sVsQ_omega regions_re_40}) and $Re=3900$. (\ref{Q_sVsQ_omega_regions_re=3900}) using  $(Q_S, Q_\Omega)$, \textcolor{red}{Red}: Boundary layer and wake regions, \textcolor{blue}{Blue}: Outer flow region.}
    \label{regions_Q_s_Q_omega}
    \end{figure}

If we exclude $R_S$ from the feature space $E$ and proceed to carry out the clustering with $(Q_S, Q_\Omega)$, we retrieve the results presented in figure~\ref{regions_Q_s_Q_omega}. Part of the wake is identified as an outer flow region in the laminar case as shown in figure~\ref{Q_sVsQ_omega regions_re_40}. For the turbulent case, the detected wake region is narrower compared to the one detected with feature space $E$ in figure~\ref{GM_3900}.
 To investigate the quality of the detected regions using $(Q_S, Q_\Omega)$, we compare the scatter plot of the invariants maps $(-Q_S, R_S)$ in the detected outer flow regions using the feature space $E$ and $(Q_S, Q_\Omega)$. 
  \begin{figure}
     \centering
      \begin{subfigure}[h!]{0.4\textwidth}
         \centering
         \includegraphics[width=\textwidth]{Results/R_sVsQ_s_Inviscid_Region_Re_40.png}
         \caption{$(R_S, -Q_S)$ scatter plot in the detected outer flow region at $Re=40$ with feature space $E$.}
         \label{Q_sVsR_s regions_re_40_E_1}
     \end{subfigure}
     \hfill
     \begin{subfigure}[h!]{0.4\textwidth}
         \centering
         \includegraphics[width=\textwidth]{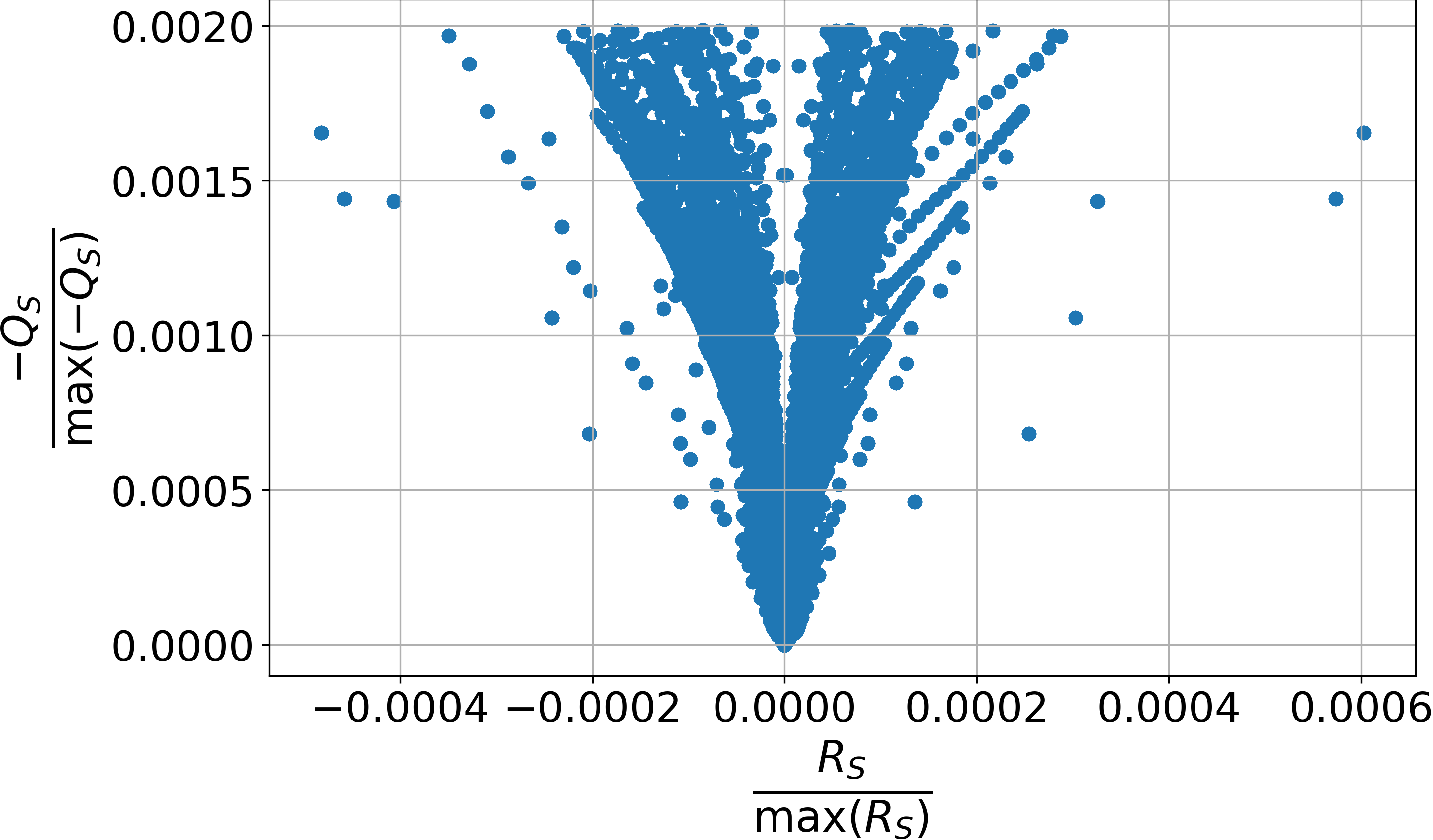}
         \caption{$(R_S, -Q_S)$ scatter plot in the detected outer flow region at $Re=40$ with $(Q_S, Q_\Omega)$.}
         \label{Q_sVsR_S_regions_re=40_Q_omega_Q_s}
     \end{subfigure}
     \begin{subfigure}[h!]{0.4\textwidth}
         \centering
         \includegraphics[width=\textwidth]{Results/Re_3900/Gaussian_mixture/Q_sVsR_s_Inviscid_Re_3900.png}
         \caption{$(R_S, -Q_S)$ scatter plot in the detected outer flow region at $Re=3900$ with feature space $E$.}
         \label{Q_sVsR_s regions_re_3900_E_1}
     \end{subfigure}
     \hfill
     \begin{subfigure}[h!]{0.4\textwidth}
         \centering
         \includegraphics[width=\textwidth]{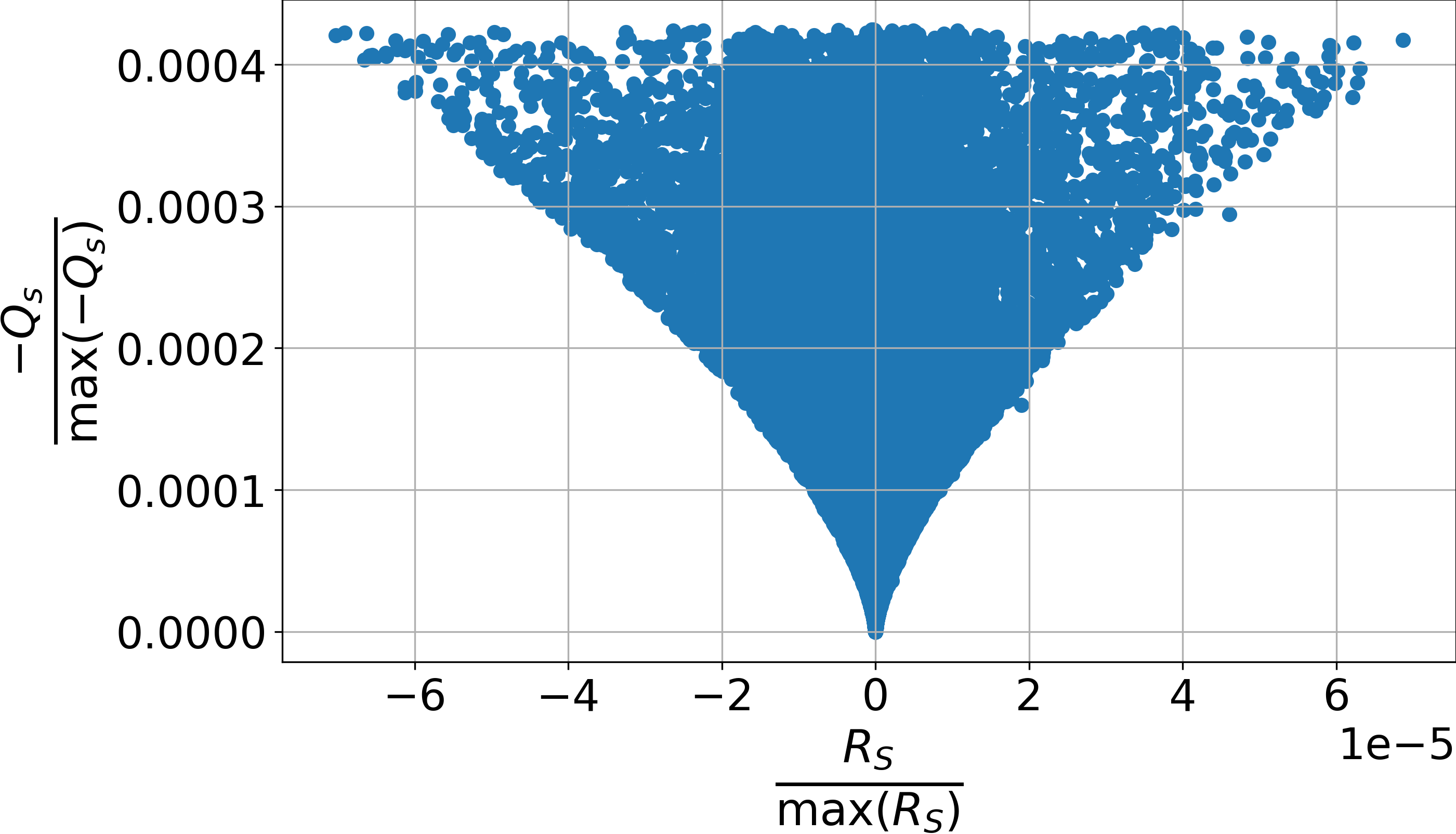}
         \caption{$(R_S, -Q_S)$ scatter plot in the detected outer flow region at $Re=3900$ with $(Q_S, Q_\Omega)$.}
         \label{Q_sVsR_s_regions_re=3900_Q_omega_Q_s}
     \end{subfigure}
     \caption{($R_S, -Q_S$) scatter plot in the detected outer flow region by GMM using the feature space $E$ and  $(Q_S, Q_\Omega)$.}
     \label{scatter_plot_Q_s_Q_omega_regions}
     \end{figure}
The results are presented in the scatter plots of figures~\ref{Q_sVsR_S_regions_re=40_Q_omega_Q_s} and \ref{Q_sVsR_s_regions_re=3900_Q_omega_Q_s}. Relatively higher strain production and destruction are detected by GMM in the outer region when using $(Q_S, Q_\Omega)$, compared to the one detected using the feature space $E$. For the case of the cylinder at $Re=3900$, the $\frac{R_S}{max(R_S)}$ values increase by a factor of $10^{2}$ as shown in figures~\ref{Q_sVsR_s regions_re_3900_E_1}~and~\ref{Q_sVsR_s_regions_re=3900_Q_omega_Q_s}.

\subsubsection{Feature space $(\boldsymbol{R_S},\boldsymbol{Q_\Omega})$}
\begin{figure}
     \begin{subfigure}[b!]{0.4\textwidth}
         \centering
     \includegraphics[width=\textwidth]{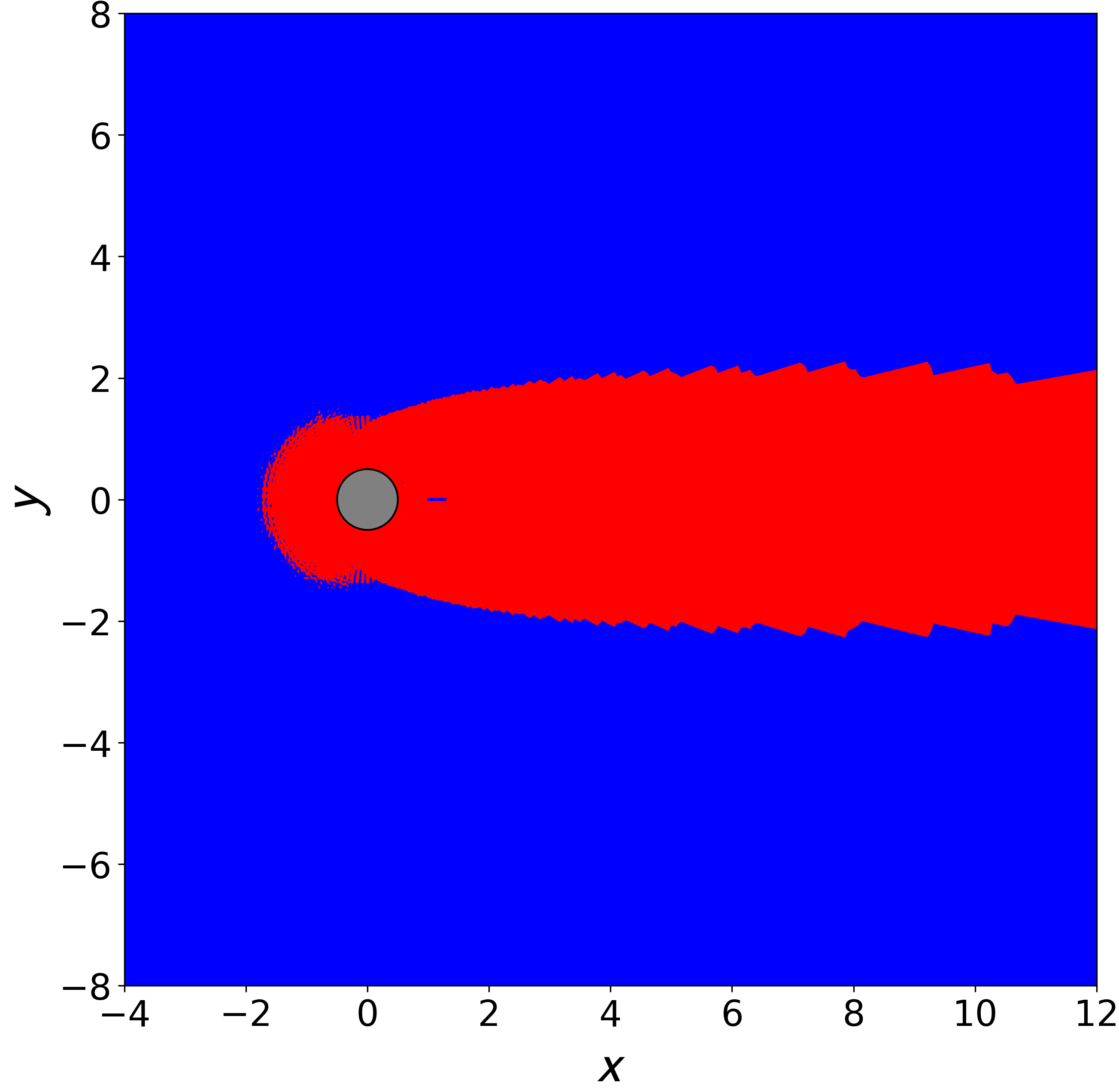}
         \caption{ }
         \label{R_sVsQ_omega regions_re_40}
     \end{subfigure}
     \hfill
     \begin{subfigure}[b!]{0.57\textwidth}
         \centering
         \includegraphics[width=\textwidth]{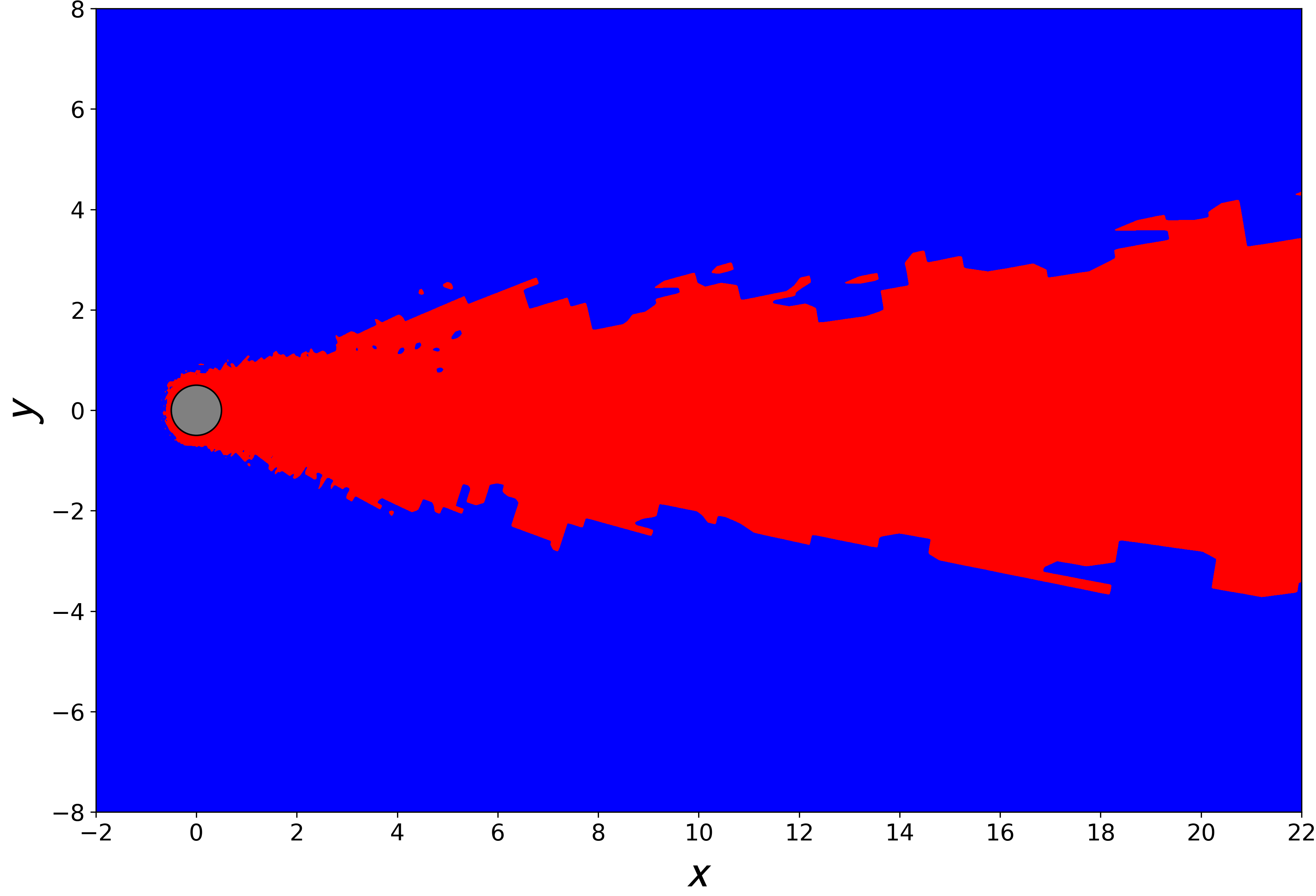}
         \caption{ }
         \label{R_sVsQ_omega_regions_re=3900}
     \end{subfigure}
     \caption{Flow regions detected by GMM for flow past a cylinder at $Re=40$. (\ref{Q_sVsQ_omega regions_re_40}) and $Re=3900$. (\ref{Q_sVsQ_omega_regions_re=3900}) using  $(R_S, Q_\Omega)$, \textcolor{red}{Red}: Boundary layer and wake regions, \textcolor{blue}{Blue}: Outer flow region.}
    \label{regions_R_s_Q_omega}
 \end{figure}

Performing the GMM clustering using $(R_S,Q_\Omega)$ leads to the results reported in figure \ref{regions_R_s_Q_omega} for both laminar and turbulent cases. 

\begin{figure}
     \centering
      \begin{subfigure}[h!]{0.4\textwidth}
         \centering
         \includegraphics[width=\textwidth]{Results/Q_omegaVsQ_s_Inviscid_Region_Re_40.png}
         \caption{$(Q_\Omega, -Q_S)$ scatter plot in the detected outer flow region at $Re=40$ with feature space $E$.}
         \label{Q_sVsQ_omega regions_re_40_E}
     \end{subfigure}
     \hfill
     \begin{subfigure}[h!]{0.4\textwidth}
         \centering
         \includegraphics[width=\textwidth]{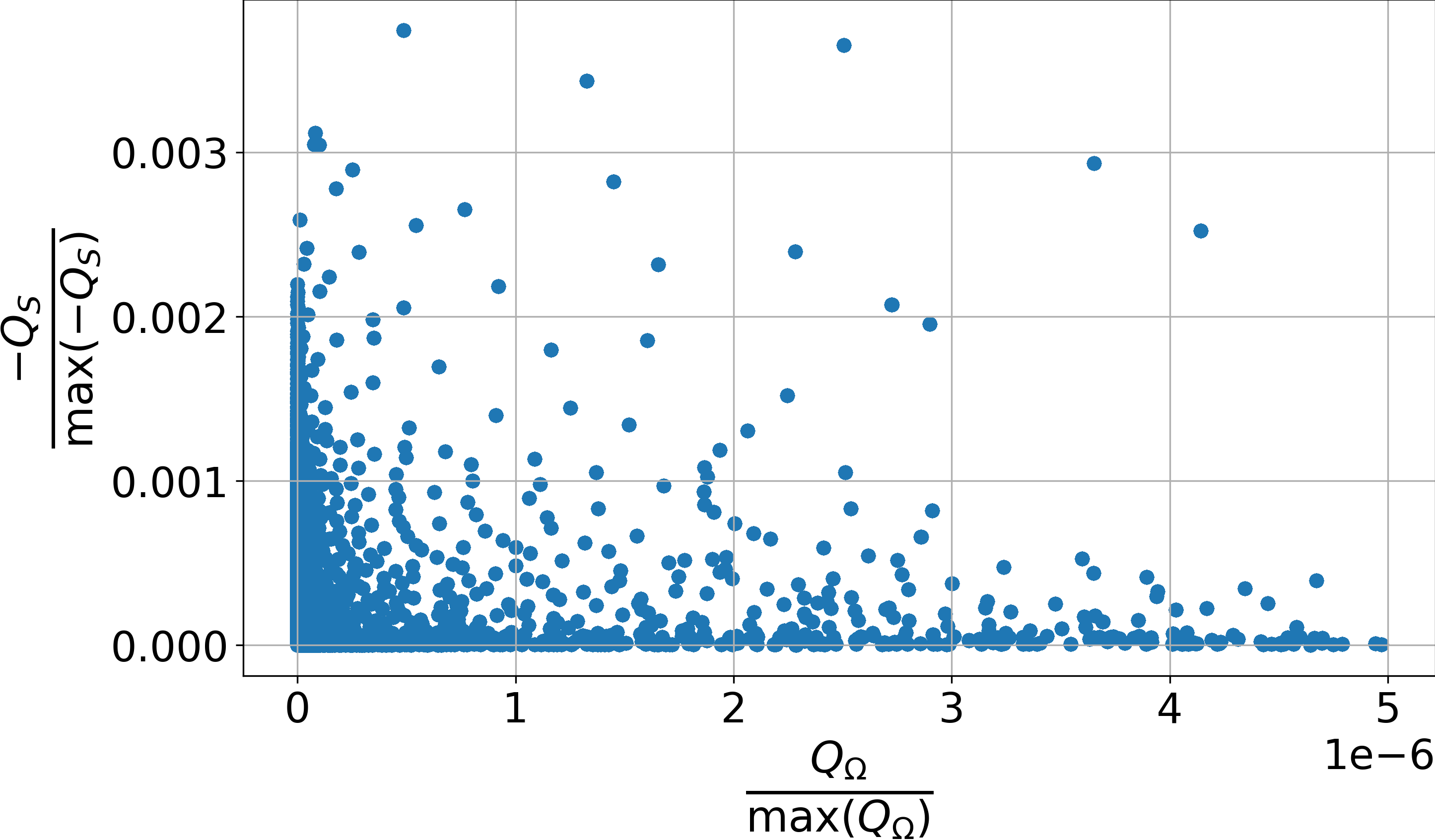}
         \caption{$(Q_\Omega, -Q_S)$ scatter plot in the detected outer flow region at $Re=40$ with $(R_S, Q_\Omega)$.}
         \label{Q_sVsQ_omega_regions_re=40_Q_omega_R_s}
     \end{subfigure}
     \begin{subfigure}[h!]{0.4\textwidth}
         \centering
         \includegraphics[width=\textwidth]{Results/Re_3900/Gaussian_mixture/Q_sVsQ_omega_Inviscid_Re_3900.png}
         \caption{$(Q_\Omega, -Q_S)$ scatter plot in the detected outer flow region at $Re=3900$ with feature space $E$.}
         \label{Q_sVsQ_omega_regions_re_3900_E}
     \end{subfigure}
     \hfill
     \begin{subfigure}[h!]{0.4\textwidth}
         \centering
         \includegraphics[width=\textwidth]{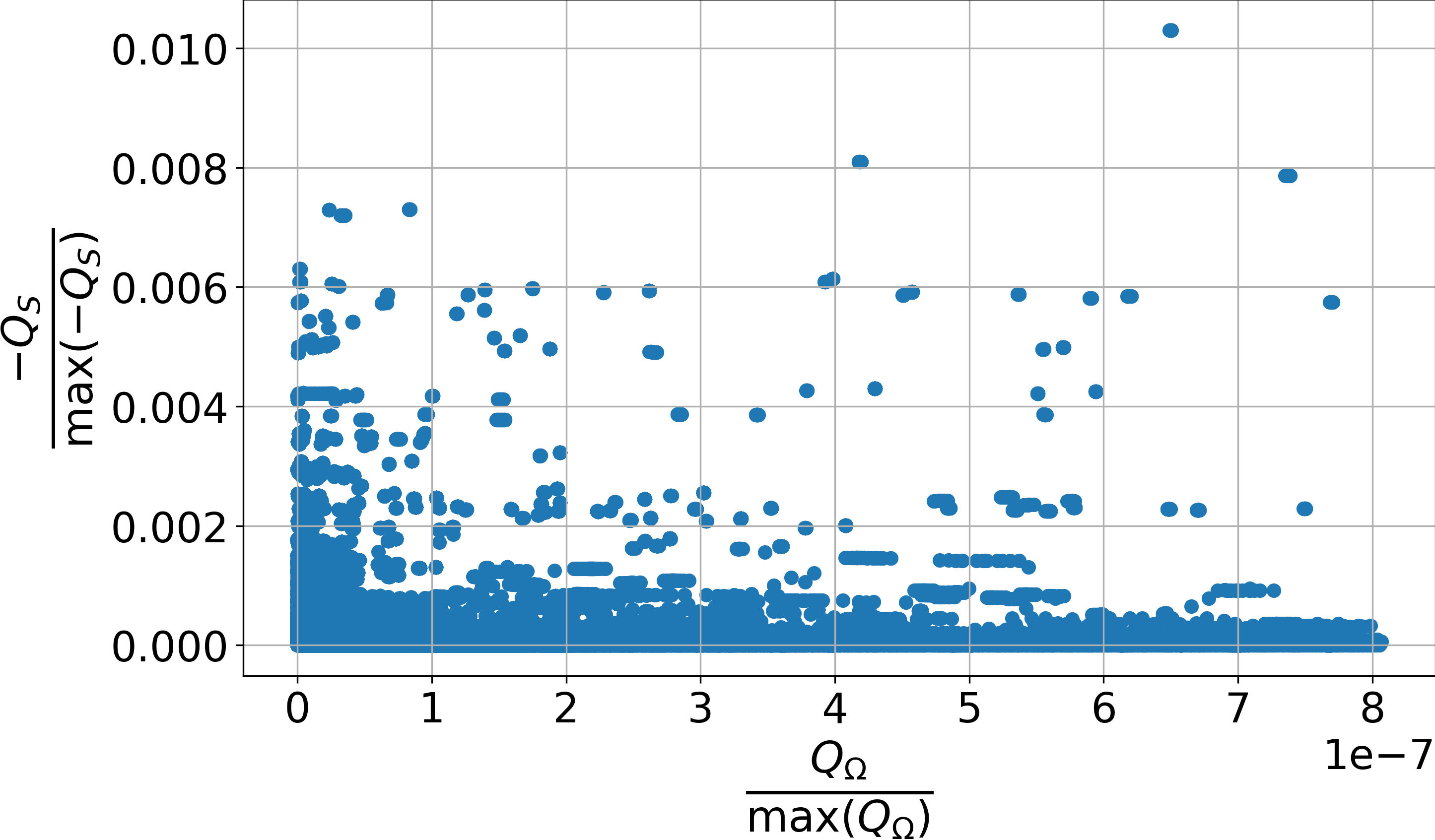}
         \caption{$(Q_\Omega,-Q_S)$ scatter plot in the detected outer flow region at $Re=3900$ with $(R_S, Q_\Omega)$.}
         \label{Q_sVsQ_omega_regions_re=3900_Q_omega_R_s}
     \end{subfigure}
     \caption{($Q_\Omega,-Q_S$) scatter plot in the detected outer flow region by GMM using the feature space $E$ and  $(R_S, Q_\Omega)$.}
     \label{scatter_plot_R_s_Q_omega_regions}
     \end{figure}

The regions obtained using $(R_S, Q_\Omega)$ in figure~\ref{regions_R_s_Q_omega} are similar to the regions obtained from the GMM with feature space $E$. However, as shown in figure~\ref{scatter_plot_R_s_Q_omega_regions}, the scatter plot of $(-Q_S, Q_\Omega)$ shows that in the detected outer flow region the value of $\frac{-Q_S}{max(-Q_S)}$ increased by a factor of $10^2$ when using $(R_S, Q_\Omega)$ to cluster the data compared to the results obtained using the feature space $E$. Similar conclusions can be observed for the turbulent test case, see figures \ref{Q_sVsQ_omega_regions_re_3900_E} and \ref{Q_sVsQ_omega_regions_re=3900_Q_omega_R_s}. 

We conclude that the proposed feature space $E=\left( Q_{S},R_{S}, Q_{\Omega} \right)$ is the most robust of all the test spaces to detect laminar and turbulent regions dominated by viscosity, vorticity and turbulence, allowing a distinction from the outer inviscid (or potential) flow. 
 
\clearpage

\section{Conclusions}\label{sec:conclusions}
We have proposed a robust clustering machine learning algorithm to distinguish viscous/turbulent dominated regions from inviscid/irrotational regions. The same feature space composed by Galilean invariants (two strain rate tensor and one rotational tensor invariant) is shown to be effective for clustering in laminar and turbulent regimes. 
Clustering can provide improved results compared to those of classic sensors. However, classic sensors require the tuning of parameters with arbitrary thresholds, whilst clustering is parameter-free.

In future work, the regions characterised as viscous/turbulent will be used for local mesh adaptation, where the resolution will be increased in the regions of interest to enhance local accuracy while reducing the computational cost.

\section*{Acknowledgements}
Gerasimos Ntoukas and Esteban Ferrer would like to thank the European Union’s Horizon 2020 Research and Innovation Program under the Marie Skłodowska-Curie grant agreement No 813605 for the ASIMIA ITN-EID project. Additionally, the authors gratefully acknowledge the Universidad Politécnica de Madrid (www.upm.es) for providing computing resources on Magerit Supercomputer.

\newpage


\end{document}